\documentclass[useAMS,usenatbib]{mn2e}
\usepackage{natbib}
\usepackage{graphicx}
\usepackage{amssymb}
\usepackage{cite}
\usepackage{threeparttable}

\usepackage{wasysym}

\title[Milky Way disc model - III]{Towards a fully consistent Milky Way disc model - III.\\Constraining the initial mass function}
\author[J. Rybizki and A. Just]{J. Rybizki$^{1,2}$\thanks{E-mail: rybizki@ari.uni-heidelberg.de, just@ari.uni-heidelberg.de} and A. Just$^{1}$\\
$^{1}$Astronomisches Rechen-Institut, Zentrum f\"{u}r Astronomie der Universit\"{a}t Heidelberg, M\"{o}nchhofstrasse 12-14, 69120 Heidelberg, Germany\\
$^{2}$Fellow of the International Max Planck Research School for Astronomy and
  Cosmic Physics at the University of Heidelberg (IMPRS-HD)}
\begin{document}

\date{Accepted 2014 December 22. Received 2014 December 18; in original form 2014 July 30}

\pagerange{\pageref{firstpage}--\pageref{lastpage}} \pubyear{2002}

\maketitle
\label{firstpage}
\begin{abstract}
We use our vertical Milky Way disc model together with Galaxia to create mock observations of stellar samples in the solar neighbourhood. We compare these to the corresponding volume complete observational samples of dereddened and binary accounted data from Hipparcos and the Catalogue of Nearby Stars.\\
Sampling the likelihood in the parameter space we determine a new fiducial initial mass function (IMF) considering constraints from dwarf and giant stars. The resulting IMF observationally backed in the range from 0.5 to 10\,M$_\odot$ is a two slope broken power law with $-1.49\pm0.08$ for the low mass slope, a break at $1.39\pm0.05$\,M$_\odot$ and a high mass slope of $-3.02\pm0.06$.\\
The Besan\c{c}on group also converging to a similar IMF even though their observational sample being quite different to ours shows that the forward modelling technique is a powerful diagnostic to test theoretical concepts like the local field star IMF. 
\end{abstract}


\section{Introduction}
The initial mass function (IMF) has seen numerous re-determinations since the seminal work of \citet{Sa55}. The most popular being broken power laws \citep{Sc86,Kr93} or lognormal with a Salpeter-like high mass slope \citep{Ch03}. Originally the IMF was derived by luminosity functions (LFs) of the solar neighbourhood via the luminosity-mass relation and stellar lifetimes. More sophisticated derivations account for the observational sample's star formation history (SFH) \citep{Sc59} and connectedly the scale height dilution \citep{Mi79}.

Another more direct approach is to look at young stellar clusters and derive an IMF from these spatially and temporally confined starburst populations. This yields a range of different IMFs \citep{Va60,Di14} which is not necessarily in conflict \citep{We05} with a global IMF being able to describe the field stars in the Milky Way disc. In this paper we try to get hold of the theoretical concept of a time independent local Milky Way IMF by exploring its effect within our disc model in the realm of observables.

The series of papers started with \citet[hereafter paper\,I]{JJ} where a local vertical Milky Way disc model was constructed using dynamical constraints of main sequence (MS) stars. The main advantage of this method is its weak dependence on the assumed IMF as only the integrated mass loss goes into the dynamical model. Other determinations as for example \citet{Ha97a} have a degeneracy of their derived SFH and their IMF's high mass slope. As a result the old Besan\c{c}on model \citep{Ro03} using the constant SFH from \citet{Ha97a} had to be revised recently \citep{Cz14} to a decreasing one in line with different determinations from chemical evolution models \citep{Ch97}, extragalactic trends \citep{Ly07} and dynamical constraints \citep{Au09,JJ}.

In paper\,II of this series \citep{Ju11} the disc model was further constrained by comparing Sloan Digital Sky Survey (SDSS) data with predicted star counts of the north galactic pole (NGP). This yielded the fiducial model A (from now on called JJ-model) with fixed SFH, a well defined local thick disc model (around 6\,\% of local stars) which showed very good agreement between our model and the NGP data being confirmed in \citet{Cz14}.

After fixing the SFH, age-velocity dispersion relation (AVR) and age-metallicity relation (AMR) we now want to further build up a consistent disc model by constraining the IMF using local star counts and taking into account turn-off stars and giants as well. In paper I the present day stellar mass function (PDMF) was already converted into an IMF considering scale height correction and finite stellar lifetimes. Binarity and reddening were not accounted for producing a very steep high mass slope with $\alpha=4.16$ which is also related to a relatively high break in the power law at $1.72\,\mathrm{M}_\odot$.

In this paper we redo the selection of volume complete samples from the revised Hipparcos Catalogue \citep{le07} and an updated version of the Catalogue of Nearby Stars\,4 \citep[which will soon be published as CNS\,5]{Ja97} also giving us a sound statistical sample to investigate the local stellar mass density (see discussion in sec\,\ref{sec:LF}). We use Galaxia \citep{Sh11} to create mock observations from the JJ-model with arbitrary IMFs. We find a new fiducial IMF describing the data by sampling the two slope broken power law parameter space using a Markov chain Monte Carlo technique. The model likelihood given the data is approximated assuming discrete Poisson probabilities for different magnitude bins differentiating between dwarfs and giants.\\
A similar determination of the IMF was done in the \citet{Cz14} Besan\c{c}on update with matching results. Instead of local luminosity functions they used the colour projections of Tycho\,2 \citep{Hoe00} data in different directions to determine their best fit. Their modelling machinery is very modular being able to incorporate different stellar evolutionary- and atmosphere models as well as accounting for binarity or reddening from the model's side. On the other hand they have so many degenerate free parameters that they do not explore the full IMF parameter space but use pre-determined ones from literature.
 
Overall the holistic approach with a concordance Galaxy model in the background able to produce mock observations to test theoretical concepts like for example the IMF is a very promising technique especially with the underlying physical models getting more and more refined with increasing observational evidence. One unsolved problem which is also pointed out by \citet{Cz14} are missing detailed 3d extinction data of the local ISM.

In section\,\ref{sec:model} we explain how to create mock observations from our vertical disc model using Galaxia. Section\,\ref{sec:obs} describes the reduction of Hipparcos and CNS\,5 data to obtain our observational sample, followed by section\,\ref{sec:stats} on the likelihood determination in the IMF parameter space. The results are then presented in section\,\ref{sec:results} together with a comparison to widely used IMFs. In section\,\ref{sec:disc} the results are put into context followed by the conclusions in section\,\ref{sec:conclusion}.
\begin{table}
\begin{center}
\begin{tabular}{ |l|l| }
  \hline
  \multicolumn{2}{|c|}{Abbreviations} \\
  \hline
  IMF & Initial Mass Function \\
  PDMF & Present Day stellar Mass Function\\
  LF & Luminosity Function\\
  VMag & Absolute V\,magnitude \\
  B-V & B-V colour\\
  CMD & Colour Magnitude Diagram\\
  MS & Main Sequence\\
  SFR & Star Formation Rate \\
  SFH & Star Formation History \\
  AVR & Age Velocity-dispersion Relation\\
  AMR & Age Metallicity Relation\\
  CNS & Catalogue of Nearby Stars\\
  paper\,I & \citet{JJ}\\
  paper\,II & \citet{Ju11}\\
  JJ-model & paper I\,\&\,II fiducial model A \\
  KTG\,93 & \citet{Kr93}\\
  Cha\,03 & \citet{Ch03}\\
  Bes\,B & Default model B of \citet{Cz14}\\
  Galaxia & Tool to synthesise observations \citep{Sh11}\\  
  FH\,06 & \citet{Fl06}\\ 
  NGP & North Galactic Pole\\
  SSP & Simple Stellar Population\\
  CEM & Chemical Enrichment Model\\  
  $\Delta$mag & Magnitude difference of stars in a binary system\\
  ISM & InterStellar Medium\\
  BD & Brown Dwarf\\
  WD & White Dwarf\\
  MCMC & Markov chain Monte Carlo\\
  PDF & Probability Distribution Function\\
  \hline
  \label{tab:abbreviations}
\end{tabular}
\end{center}
\end{table}
\section{Synthesising a local disc model}
\label{sec:model}
In paper\,I a self-consistent dynamical vertical Milky Way disc model was developed. Combined with constraints from star counts of the NGP in paper\,II a fiducial disc model (JJ-model) was chosen. The JJ-model fixes the SFH, the AVR, a simple metal enrichment law and from that predicts the stellar vertical disc structure in terms of kinematics, star counts, ages and metallicities as functions of distance to the Galactic plane.\\
In order to obtain mock observations of the solar neighbourhood we turn our vertical disc model into a local representation used as an input for Galaxia to synthesise stellar samples.    
\subsection{The disc model locally}
\begin{figure}
\centering
\includegraphics[width=0.49\textwidth]{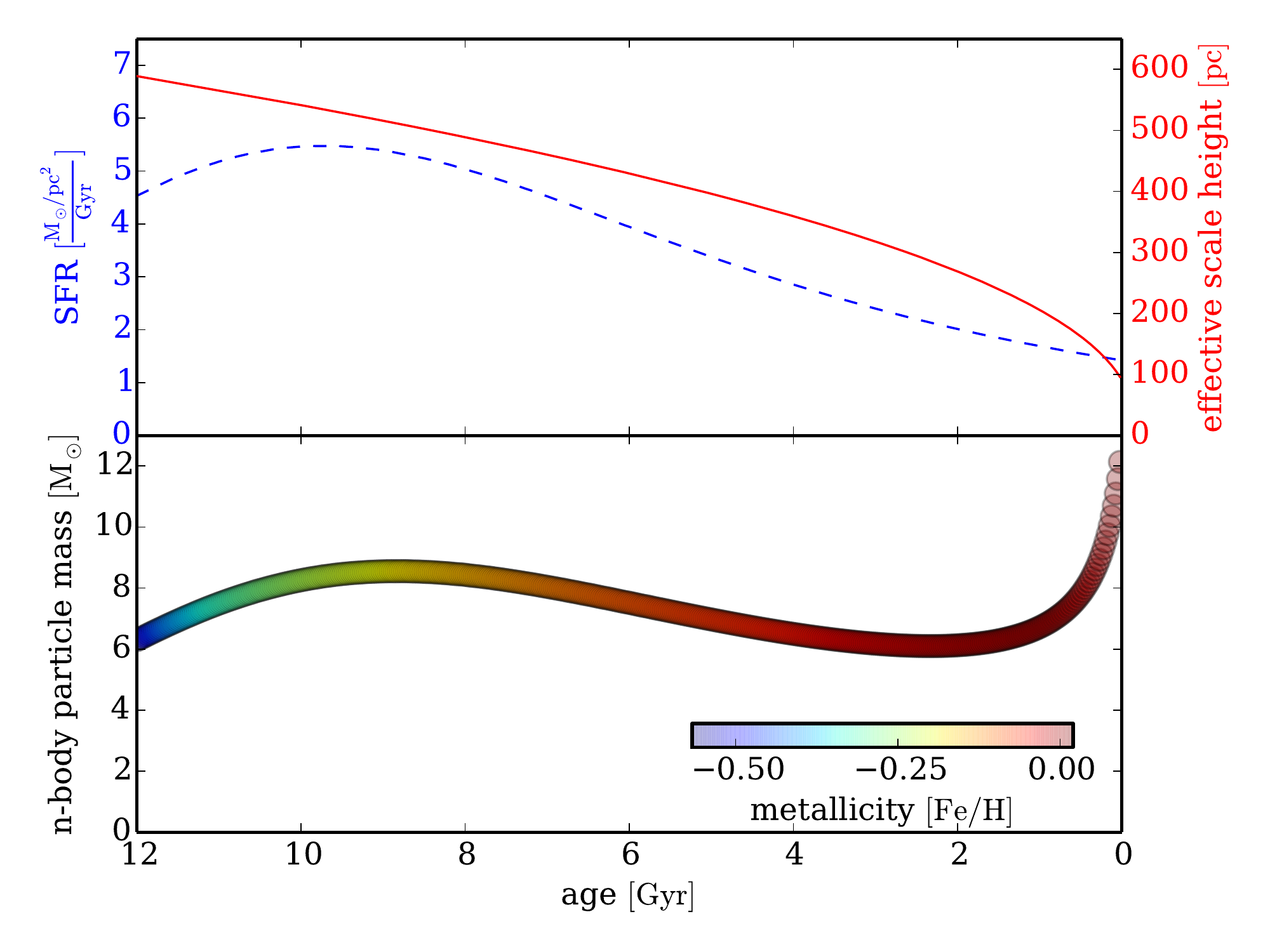}
\caption[CSP]{The upper panel shows in dashed blue the global SFH and in solid red the effective scale height of the JJ-model (at solar galactocentric distance).

In the lower panel the input n-body data which is passed to Galaxia to create the 25\,pc sphere is displayed. The age, mass and metallicity of each SSP are visualised.}
\label{fig:SFH}
\end{figure}
As we want to compare our JJ-model to volume complete stellar samples in the solar neighbourhood we construct 7\,spheres that are disjunct in absolute V\,magnitude and have heliocentric distances from 20 up to 200\,pc (see table\,\ref{tab:hipparcos_dwarf}\,\&\,\ref{tab:hipparcos_giant} for detailed limits). In the following we explain how we prepare n-body particle representations of our model for each sphere such that Galaxia can turn them into mock observations. 

As our model is evolved over 12\,Gyr in 25\,Myr steps we construct 480\,n-body particles (plus one for the thick disc) with specific ages, masses and metallicities for each sphere separately. These are passed to Galaxia where each particle is turned into a simple stellar population (SSP) according to a prescribed IMF. The two youngest SSPs are split up age-wise so that the youngest stars coming from Galaxia are 6.25\,Myr old. We assign arbitrary phase-space information to the particles since the final observable is a local colour magnitude diagram (CMD) based on volume complete subsamples.

We will now illustrate the construction of a local n-body representation of our model for the 25\,pc sphere following figure\,\ref{fig:SFH}:\\
The upper panel depicts the global SFH in units of surface density per time and the effective scale height, $\mathrm{h}_\mathrm{eff}$, over age for the JJ-model. From that a local mass density, $\rho_0$, for each SSP can be calculated which then is multiplied with the volume of the 25\,pc sphere, $\mathrm{V_{25}}$, to obtain
\begin{equation}
\mathrm{M}_{25\,\mathrm{pc}}(t_i) = \mathrm{V}_{25\,\mathrm{pc}}\rho_0(t_i) = \frac{4}{3}\pi \mathrm{r}_{_{25\,\mathrm{pc}}}^3\frac{\mathrm{SFH}(t_i)\cdot\,0.025\,\mathrm{Gyr}}{2\,\mathrm{h}_\mathrm{eff}(t_i)}
\end{equation}
which is shown in the lower panel of figure\,\ref{fig:SFH} as resulting n-body particle masses. Colour-coded the chemical enrichment law (AMR) is depicted which is also coming from the JJ-model as an analytic function but with an added Gaussian scatter of 0.13\,dex standard deviation (tested with GCS data, cf. fig. 15 of paper\,I).

For the thin disc 3552\,M$_\odot$ ($\mathcal{M}_\mathrm{IMF,thin,25\,pc}$) in particle mass is passed to Galaxia for the 25\,pc sphere
\begin{equation}
\mathcal{M}_\mathrm{IMF,thin,25\,pc}=\int\limits_{0\,\mathrm{Gyr}}^{12\,\mathrm{Gyr}}\mathrm{M}_{25\,\mathrm{pc}}(t)\mathrm{d}t\simeq\sum_{i=1}^{480}\mathrm{M}_{25\,\mathrm{pc}}(t_i).
\end{equation}
This is by construction of the JJ-model the gas mass that was used to create the thin disc stars (and in the meanwhile also remnants) still residing in the 25\,pc sphere. Nowadays only a fraction of that is left in stars due to stellar evolution ($\mathcal{M}_\mathrm{PDMF,25\,pc}$, cf. section\,\ref{sec:LF}). In the mass-age distribution of the n-body data the peak from the global SFH (dashed blue line) around 10\,Gyr can still be recognised. The increase for younger stellar populations stems from the decreasing effective scale height confining these stars closer to the Galactic plane (i.e. a bigger fraction of them is found in the local sphere, cf. fig. 14 of paper\,I).

The thick disc is implemented by inserting a single starburst (i.e. one SSP and equivalently one n-body particle) with 6.5\,\% of the thin disc mass 
\begin{equation}
\mathcal{M}_\mathrm{IMF,discs}=\mathcal{M}_\mathrm{IMF,thin}+\mathcal{M}_\mathrm{IMF,thick}=1.065\cdot\mathcal{M}_\mathrm{IMF,thin}
\end{equation}
12\,Gyr ago with a metallicity of $[\mathrm{Fe}/\mathrm{H}]= -0.7$ resulting in a present day thick disc mass fraction of around 5\,\%. Since the local density of the stellar halo is negligible compared to the disc we do not consider it as a separate component in this work. 

A change in the IMF will affect the mass fraction remaining in the stellar component (cf. sec. 2.5 paper\,I) so we introduce a mass factor ($\mathrm{mf}$) scaling our model's total mass ($\mathcal{M}_\mathrm{IMF}$) to fit the observed stellar mass density
\begin{equation}
\mathcal{M}_\mathrm{IMF}=\mathrm{mf}\cdot\mathcal{M}_\mathrm{IMF,discs}.
\label{eq:totalmass}
\end{equation}
When going to larger heights above the Galactic plane we also need to correct for decreasing vertical density profiles. For the 25\,pc sphere the deviation from homogeneous density distribution is still negligible but this changes with larger spheres and especially with young stellar populations. For example when rescaling the star count in the highest magnitude bin in table\,\ref{tab:hipparcos_dwarf} with a mean age of 0.1\,Gyr the number must be increased by 54\,\% to account for the low scale height of these stars (cf. figure\,2 in paper\,II).
\subsection{Mock observations with Galaxia}
\label{galaxia}
Galaxia \citep{Sh11} is a tool to generate mock catalogues from either analytic models or n-body data. It already has a default Besan\c{c}on-like model \citep{Ro03} but the updates \citep{Ro12,Cz14} are not implemented yet.

In the previous subsection we constructed particles representing our model locally which we now pass for each sphere separately to Galaxia together with the disjunct VMag limits building up a CMD successively. 

So far Galaxia uses Padova isochrones \citep{Ma08} which have problems reproducing the lower end of the MS and the red clump. We include their revised templates (PARSEC version 1.2\,S\footnote{http://stev.oapd.inaf.it/cgi-bin/cmd}) where only minor differences at low mass stars persist (cf. lower MS in figure\,\ref{fig:division}). This remaining discrepancy is also pointed out in the release paper \citet[fig.\,A3]{Ch14} but should have negligible effect on the star counts in our used magnitude range.

Binaries, white dwarfs or other remnants are not implemented in Galaxia yet but an update is being planned (private communication, Sharma 2015). When inspecting the CMD in figure\,\ref{fig:division} a second blue-shifted MS in the synthesised catalogue is visible which comes from our distinct thick disc metallicity.

Beside being able to change the IMF according to which Galaxia is distributing the particle masses into stars we are using it as a black box. Specifying a photometric system will already yield a detailed stellar catalogue\footnote{See http://galaxia.sourceforge.net/Galaxia3pub.html for detailed instructions} in terms of a random realisation of our local model representation as for example depicted in figure\,\ref{fig:division} for our newly determined IMF.
\begin{table*}
\centering
\caption{Observational sample and mock catalogues - dwarf stars}
\begin{tabular}{c|c c c c c c c c c c c}
\hline
Catalogue&d & M$_V$-limits & N$_{fin}$ & $\frac{\sigma_\pi}{\pi} >15 $\,\% & CNS\,5 & N$_{25}$ & JJ$_{25}$ & log-likelihood & Mean mass & Mean age \\
&(pc) & (Mag) & \# & \# lost & 25\,pc &\multicolumn{2}{|c|}{rescaled$^\dagger$ to 25\,pc}
& $\ln\left(\mathcal{L}/\mathrm{P}_\mathrm{max}\right)$ & M$_\odot$ & Gyr \\
\hline
&200 & $],-1.5]$    & 98  & 0  & 0    & 0.28      & 0.13    	& -21.9	& 6.4   & 0.1\\
&200 & $[-1.5,-0.5]$  & 233   & 3  & 1    & 0.62      & 0.63    & \phantom{-}0.00	& 4.0   & 0.2 \\
&200 & $[-0.5,0.5]$   & 901   & 12 & 4    & 2.26      & 2.60    & -9.20	& 2.9   & 0.3 \\
Hipparcos&100 & $[0.5,1.5]$& 520& 2  & 15   & 8.61      & 9.09  & -0.76	& 2.2   & 0.5 \\
&75 & $[1.5,2.5]$     & 677   & 1  & 27   & 25.6      & 24.4    & -0.81	& 1.7   & 1.0 \\
&50 & $[2.5,3.5]$     & 518   & 1  & 62   & 65.1      & 59.9    & -1.79	& 1.4   & 2.3 \\
&30 & $[3.5,4.5]$     & 200   & 0  & 110  & 115.7     & 146.5   & -5.90	& 1.1   & 5.3 \\
\hline
&25 & $[4.5,5.5]$     & 191   & 4  & 191  & 191       & 190.3   & \phantom{-}0.00	& 0.9   & 6.1 \\
&25 & $[5.5,6.5]$     & 198   & 11 & 198  & 198     & 207.9   	& -0.22	& 0.8   & 6.4 \\

CNS\,5&25 & $[6.5,7.5]$ & 193   & 16 & 193  & 193       & 196.0 & -0.02		& 0.7   & 6.5 \\
&25 & $[7.5,8.5]$     & 207   & 13 & 207  & 207     & 205.3   	& -0.01		& 0.6   & 6.6 \\
&20 & $[8.5,9.5]$     & 139   & 15 & \phantom{$^\star$}245$^\star$ & 271.5  & 258.0 & -0.20 & 0.5   & 6.7 \\
\hline\hline
&$\Sigma$ & $],9.5]$  & 4075  & 78 & 1253 & 1278.8  & 1300.8  	& -40.8	& 0.8 & 6.0 \\

\end{tabular}
\begin{tablenotes}
\item \textit{Catalogue} shows from which source the observational sample is drawn, \textit{d} gives the heliocentric distance of stars included (sphere size), \textit{M$_V$-limits} gives the magnitude range of each bin, \textit{Nfin} 
is the final star count in each bin, \textit{$\frac{\sigma_\pi}{\pi} >15$\,\%} is the number of stars thrown out due to high distance errors, \textit{CNS\,5} gives the number of stars within the volume complete 25\,pc sphere, the next two columns give the star counts of the observations and our JJ-model rescaled to 25\,pc. The JJ-model with newly determined IMF is averaged over 400 random realisations, \textit{log-likelihood} shows the probability of each bin after equation\,\ref{eq:chi} normed with the maximal possible probability (cf. section\,\ref{sec:likelihood}) in natural logarithm which indicates each bin's impact on the likelihood function (we will call it {\it penalty} because our likelihood can be seen as a reward function for the MCMC simulation), the \textit{mean mass} and \textit{mean age} show the values for the corresponding JJ magnitude bins where the sum at the bottom is an average of all stars within 25\,pc.
\item $^\star$for this magnitude bin volume completeness is not given so the 139\,stars from the 20\,pc sphere have been rescaled to 25\,pc yielding 271.5\,stars which is 10\,\% more than the 245\,stars observed in the 25\,pc sphere
\item $^\dagger$rescaling the volume and accounting for the density profile of the magnitude bin's mean age population
\end{tablenotes}
\label{tab:hipparcos_dwarf}
\end{table*}

\begin{table*}
\centering
\caption{Observational sample and mock catalogues - giant stars}
\begin{tabular}{c|c|c|c|c|c|c|c|c|c|c|c}
\hline
Catalogue&d & M$_V$-limits & N$_{fin}$ & $\frac{\sigma_\pi}{\pi} >15 $\,\% &CNS\,5 & N$_{25}$ & JJ$_{25}$ & log-likelihood & Mean mass & Mean age \\
&(pc) & (Mag) & \# & \# lost & 25\,pc &\multicolumn{2}{|c|}{rescaled$^\dagger$ to 25\,pc}
& $\ln\left(\mathcal{L}/\mathrm{P}_\mathrm{max}\right)$ & M$_\odot$ & Gyr \\
\hline
&200 & $],-1.5]$    & 74  & 1  & 0  & 0.16  & 0.14      	& -0.87	& 3.4 & 1.6\\
&200  & $[-1.5,-0.5]$ & 375   & 2  & 1  & 0.77  & 0.84      & -1.47	& 1.8 & 4.4 \\
&200 & $[-0.5,0.5]$   & 1341  & 20 & 3  & 2.78  & 2.29      & -23.4	& 1.7 & 3.9 \\
Hipparcos &100 & $[0.5,1.5]$ &526& 0 & 6  & 8.33  & 9.27    & -3.03	& 1.5 & 4.8 \\
&75 & $[1.5,2.5]$     & 126   & 0  & 5  & 4.70  & 3.76      & -2.99	& 1.2 & 6.3 \\
&50 & $[2.5,3.5]$     & 62  & 0  & 7  & 7.77  & 11.6      	& -5.55	& 1.2 & 6.6 \\
&30 & $[3.5,4.5]$     & 9   & 0  & 6  & 5.21  & 7.11      	& -0.34	& 1.0 & 8.8\\
\hline\hline
&$\Sigma$& $],9.5]$   & 2513  & 23 & 28 & 29.7    & 35.0    & -37.7	& 1.3 & 6.3\\
\end{tabular}
\label{tab:hipparcos_giant}
\end{table*}
\section{Observations}
\label{sec:obs}
The anchoring point for every Galaxy model in terms of observational constraints is the stellar distribution in the solar neighbourhood since detailed and volume complete samples can only be obtained here. After using the vertical component of the velocity distribution of MS stars in paper I (dynamical constraint) and the NGP star counts in paper II (vertical density distribution constraint) we now use absolute local stellar densities for dwarf and giant stars.\\ 
This is achieved by constructing different samples combining absolute magnitude cuts and heliocentric distances such that the selected stars represent a volume complete sphere. At the bright end we go up to 200\,pc in order to obtain enough massive stars and giants to have reliable statistics. Our observational sample consists of stars from the extended Hipparcos catalogue \citep{an12} and the Catalogue of Nearby Stars (CNS). CNS\,5 which we use here is an updated version of CNS\,4 \citep{Ja97} and will get published soon.\\
Fundamentally we would like to implement all observational biases on the models side and compare the synthesised mock observations to unaltered observables. In this respect the updated Besan\c{c}on model \citep{Cz14} has pushed the link between model and data in the right direction by implementing extinction models and a scheme for binary systems into their model. As Galaxia is not able to account for binaries yet and we are not providing positional information with the n-body particles we have to treat binaries and dereddening from the observational side. 
\subsection{Hipparcos}
We use the extended Hipparcos compilation (117955 entries) which cross-matches the original stars from the revised Hipparcos catalogue \citep{le07} with a large selection of different catalogues.\\
By using heliocentric distance and absolute magnitude cuts similar to paper\,I \citep[table 1]{JJ} we obtain volume complete observational spheres for different stellar magnitudes going down to 4.5\,VMag. The only further selection criteria to eliminate miss-identifications is a distance error below 15\% which reduces the sample insignificantly as visible in table\,\ref{tab:hipparcos_dwarf}\,\&\,\ref{tab:hipparcos_giant}.\\
Before the VMag and distance cuts are applied all stars which have both a binary flag and a $\Delta\mathrm{mag}$ entry in the original Hipparcos catalogue \citep{Hi97} are split up into two components. The binary correction changes the number of stars in each distance bin slightly as can be seen in table\,\ref{tab:corrections}.\\
\begin{table}
\begin{center}
\caption{Effect of binarity and dereddening on the star counts}
\begin{tabular*}{0.48\textwidth}{ c|c c c c c }
  Radius of sphere [pc] & 30 & 50 & 75 & 100 & 200 \\
  Magnitude bin [VMag] & 4th	& 3rd	& 2nd	& 1st & $<0.5$	\\
\hline  
  No correction & 204 & 591 & 793 & 1052 & 2798 \\
  Binary correction & 209 & 580 & 803 & 1046 & 2756 \\
  Extinction correction & 204 & 591 & 793 & 1053 & 3060 \\
\hline\hline
  Both corrections (N$_{fin}$) & 209 & 580 & 803 & 1046 & 3022 \\
\end{tabular*}
\label{tab:corrections}
\end{center}
\end{table}
With the binary correction we have 5394 stars in the Hipparcos sample of which 552 entries come from split up binary systems.
On average the bins lose stars when correcting for binaries because the single components get fainter than the system bringing these stars below the magnitude limit. As the fainter mag bins have smaller limiting radii only some {\it lost} stars fall into the fainter magnitude bin.\\
On scales of the investigated volume the ISM is in-homogeneously distributed with Ophiuchus and Taurus molecular cloud being the biggest absorbers in the 200\,pc sphere \citep{Sc14}. 3d extinction maps with good enough resolution are getting published \citep{La14} but the data are not available yet.
In order to deredden our stars we adopt an analytic model from \citet{Ve98} describing a homogeneous extinction depending on the distance and Galactic latitude. Due to the local bubble we set the extinction to 0 below 70\,pc distance and above 52$^\circ$ Galactic latitude (cf. \citet[fig. 4, 11]{Ve98}).

\citet{Au09} also adopted this model but only within 40\,pc of the Galactic mid-plane resulting in 4\,\% less stars in the 200\,pc sphere compared to our adaptation of the extinction model.
 We use the \citet[p.548]{Ve98} cosecant law:
\begin{equation}
E\left(B - V\right)=
\left\{ \begin{array}{l}
 0, \\
 \mathrm{E}_0(d-\mathrm{d}_0), \\
 \mathrm{E}_0\left(\frac{\mathrm{h}_0}{|\sin(b)|}-\mathrm{d}_0\right),
 \end{array} \mbox{\ if \ } 
 \begin{array}{l}
 d < \mathrm{d}_0\,\vee\, b > 52^\circ\\
 d < \frac{\mathrm{h}_0}{|\sin(b)|}\\
 d > \frac{\mathrm{h}_0}{|\sin(b)|}\end{array} \right.
\label{KTG}
\end{equation}

with $\mathrm{h}_0=55\,\mathrm{pc}$, $\mathrm{d}_0=70\,\mathrm{pc}$, $\mathrm{E}_0=4.7\times10^{-4}\,\mathrm{mag/pc}$ and $d,b$ being the heliocentric distance and the Galactic latitude. To transform from $E\left(B - V\right)$ to $A\left(V\right)$ we adopt $\mathrm{R}_0=3.1$.\\
In essence the dereddening leaves the closer samples unaltered only increasing the 200\,pc sample star count by around 10\,\%.
This is due to stars with heliocentric distance between 100 and 200\,pc that are slightly too faint without dereddening become a bit brighter and thereby satisfy the magnitude limits.\\
As the ISM is highly inhomogeneous adopting an analytic model is only a crude approximation but the best we can do at the moment and we are curious to redo the analysis with upcoming 3d extinction maps.
\subsection{Catalogue of Nearby Stars 5}
The Hipparcos sample is supplemented with stars from CNS\,5 (7251 entries) at the faint end (4.5 - 9.5\,VMag) which is a volume complete catalogue for stars brighter than 8.5 (9.5) VMag up to a distance of 25 (20)\,pc.\\
The distance of the stars is calculated from parallaxes (when advisable the photometric were preferred to the trigonometric) with a correction for the Lutz-Kelker bias from \citet[eq. 1]{an12}.\\
The 20\,pc sample includes stars from 8.5 to 9.5 VMag. After excluding 2\,WDs and 15\,stars with relative distance errors above 15\% we are left with 139 stars. For the 25\,pc sphere the absolute V\,magnitude ranges from 4.5 to 8.5. Here 44 stars have a too high error so that 789 stars remain in the final sample.\\
All together 928 stars come from the CNS\,5 of which 372 are in resolved multiple stellar systems (24 of the remaining 556 {\it single systems} are detected spectroscopic binaries). Of the 372 stars in multiple stellar systems 117 mostly primary and 87 mostly secondary components share a joint B-V value but have individual VMag entries so that they can be corrected by putting them on the MS. The MS was empirically assigned from MS stars with low parallax error.

In the end three stars reside in the CNS\,5 sample as well as in the Hipparcos sample because of slightly different VMag values directly at the VMag borders of 4.5 VMag. Those stars are excluded from the Hipparcos sample so that in our joint sample every star has a unique entry.\\
As a comparison to the Hipparcos sample we also list the 247 volume complete (within 25\,pc) CNS\,5 stars which are brighter than 4.5\,VMag in table\,\ref{tab:hipparcos_dwarf} and\,\ref{tab:hipparcos_giant}. The only peculiarity is the 2\,$\sigma$ outlier with 15 against 8.65 expected stars in the 100\,pc bin of the dwarf sample. Still the CNS\,5 sample can be seen as a valid random realisation of the enlarged Hipparcos sample.
\section{Statistical analysis}
\label{sec:stats}
 \begin{figure}
 \includegraphics[width=0.5\textwidth, page=1]{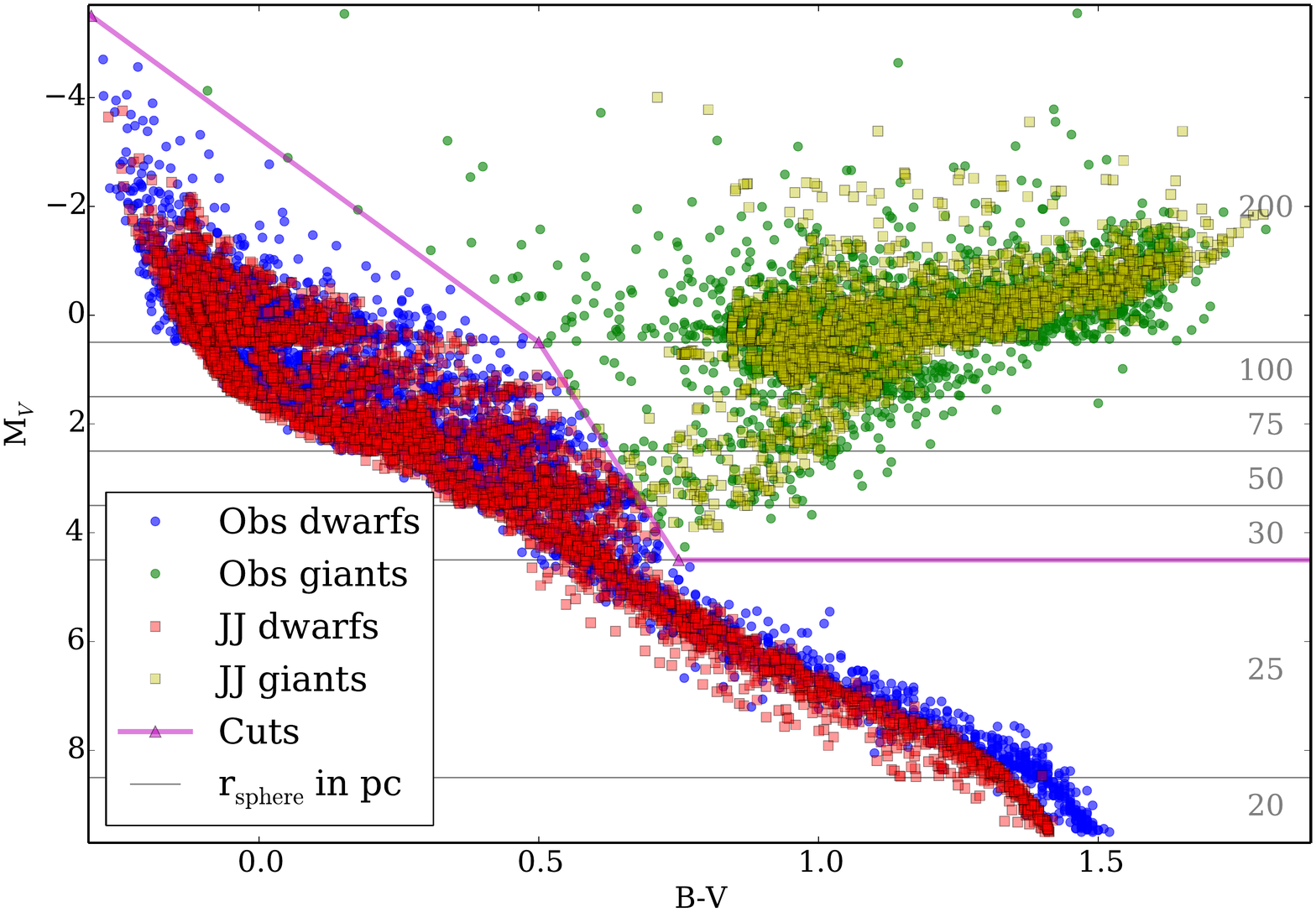}
 \caption[cuts]{CMD of the observational sample (Hipparcos stellar systems are not split up) and one random realisation from our model with the newly determined IMF. The cuts with connected triangles at (B-V,M$_\mathrm{V}$) = (-0.3,-5.5), (0.5,0.5), (0.75,4.5) and (3,4.5) for the division into dwarf and giant sample are indicated in magenta. In the right the different sphere radii are written in units of pc.}
 \label{fig:division}
 \end{figure}
For our analysis we divide the derived CMDs into dwarf and giant stars with the cuts specified in figure\,\ref{fig:division}. Since MS stars have a tight correlation between luminosity and stellar mass the dwarf sample contains information on the PDMF. The giant sample adds constraints for the integrated SFH and the IMF of higher mass stars (above 0.9 M$_\odot$).\\
To compare our observable (the LF) to a theoretical IMF we feed local representations of our model together with different IMFs to Galaxia leaving us with synthesised star counts from which we construct a likelihood assuming discrete Poisson processes (section\,\ref{sec:likelihood}). This is implemented into a Markov chain Monte Carlo (MCMC) scheme to obtain a representation of the probability distribution function (PDF) in the two slope IMF parameter space (section\,\ref{sec:bestfit}).
\subsection{Likelihood calculation}
\label{sec:likelihood}
We approximate the likelihood of our model given the data by dividing each CMD into 12\,magnitude bins for the dwarf sample and 7 for the giants (see table\,\ref{tab:hipparcos_dwarf}\,\&\,\ref{tab:hipparcos_giant}) and calculate the discrete Poisson probability distribution. The expected value is coming from our model ($m_i$) and the number of occurrences is the star count observed in each bin ($d_i$) leading to the likelihood
\begin{equation}
\mathcal{L}_\mathrm{total}=\prod\limits_{i=1}^{12+7}\mathcal{L}_i,
\mbox{\ where \ }\mathcal{L}_i=\frac{m_i^{d_i}e^{-m_i}}{d_i!}.
\label{eq:chi}
\end{equation}
The log-likelihood then follows as:
\begin{equation}
\log\mathcal{L}_\mathrm{total}=\sum\limits_{i=1}^{12+7}\big(d_i\log(m_i)-\log(d_i!)-m_i\big).
\end{equation}
For the calculation of the log factorial a very accurate approximation for $n>0$ from \citet[p. 339]{Ai88} is used:
\begin{equation}
\log n! \approx n\log n - n + \frac{\log\Big(n\big(1+4n(1+2n)\big)\Big)}{6}+\frac{\log(\pi)}{2}.
\end{equation}
We will normalise the log-likelihood with its maximal possible value, P$_\mathrm{max}=-68.5$, occurring when the observed sample is tested with itself.\\
It should be kept in mind that a linear increase in star counts results in exponentially increasing penalties for our likelihood function when the deviation in percent stays the same. For example if the expected value is 10 and the number of occurrences is 9 then we have a $\ln\left(\mathcal{L}/\mathrm{P}_\mathrm{max}\right)$ of $-0.04$. For 100 expected stars and 90 occurrences $\ln\left(\mathcal{L}/\mathrm{P}_\mathrm{max}\right)$ equals $-0.46$, for 1000 and 900 it is $-5.1$ and so forth. This on the one hand takes into account that bins with a lot of stars get a higher statistical weight but could also be dangerous when small errors in rescaling (which could come from a bad AVR or connectedly isochrones indicating wrong ages) result in large penalties for the likelihood potentially pointing our MCMC simulation to a biased equilibrium IMF parameter configuration.
\subsection{Sampling the likelihood distribution}
\label{sec:bestfit}
The variability of the outcome of Galaxia is twofold. First the IMF parameters can be varied changing the laws according to which the mock observations are produced. Second for fixed parameters the random seed of Galaxia can be changed yielding different random realisations. The latter can be minimised by averaging. We oversample each point in parameter space 400\,times so that this noise is reduced by a factor of 20 (cf. tab\,\ref{tab:bootstrap}) and should be a second order effect compared to the Poisson noise in the data (if we assume the observed stars have been randomly realised from an underlying probability distribution).\\

To sample the PDF of our parameter space we use a Python implementation \citep{Fa13} of an affine invariant ensemble sampler for MCMC \citep{Go10} where step proposals using the information of multiple walkers reduce the autocorrelation time significantly.\\
Since the overall mass turned into stars $(\mathcal{M}_\mathrm{IMF,discs})$ could be slightly different to the JJ-model we add the mass factor $(\mathrm{mf})$ as a fourth free parameter beside the three two slope IMF parameters, low mass index $(\alpha_1)$, high mass index $(\alpha_2)$, and the power law break $(\mathrm{m}_1)$:\\
\begin{equation}
\frac{\mathrm{d}n}{\mathrm{d}m}=k_\alpha m^{-\alpha}
\left\{ \begin{array}{l}
 \alpha = \alpha_1, \\
 \alpha = \alpha_2,
 \end{array} \mbox{\ if \ } \begin{array}{l}\mathrm{m}_\mathrm{low}<m<\mathrm{m}_1\\ \mathrm{m}_1<m<\mathrm{m}_\mathrm{up}\end{array} \right.
\label{eq:func}
\end{equation}
with the lower and upper mass limit of the IMF $\mathrm{m}_\mathrm{low}=0.08\,\mathrm{M}_\odot$ and $\mathrm{m}_\mathrm{up}=100\,\mathrm{M}_\odot$ being fixed and $k_\alpha$ normalising the IMF to be a continuous function that represents the mass turned into stars
\begin{equation}
\mathcal{M}_\mathrm{IMF} = \mathrm{mf}\cdot\mathcal{M}_\mathrm{IMF,discs} = \int_{\mathrm{m}_\mathrm{low}}^{\mathrm{m}_\mathrm{up}} k_\alpha m^{-\alpha+1}\mathrm{d}m.
\label{eq:mass_normalisation}
\end{equation}
For every set of parameters we use the product of the likelihoods from dwarf and giant sample
\begin{equation}
\log\mathcal{L}_\mathrm{total} = \log\mathcal{L}_\mathrm{dwarf} + \log\mathcal{L}_\mathrm{giant}
\end{equation}
to sample the parameter space. In this way the probability for each bin (12 from the dwarf sample and 7 from the giant sample) being an independent discrete Poisson process is weighted equally into the final likelihood ($\mathcal{L}_\mathrm{total}$).\\
We also investigated three slope IMFs but the gain in likelihood is small. Usually a second break in the IMF is introduced to fit the low mass regime \citep[$m < 0.5\,\mathrm{M}_\odot$]{Kr93}. 
With our observational evidence ranging from 0.5 to 10\,M$_\odot$ a second break in the IMF below this limit can not be tested. But an empirical extension of our 2 slope IMF to fulfil local mass density constraints for low mass stars will be introduced in section\,\ref{sec:empirical_fix}.
\section{Results}
\label{sec:results}
Here we present the newly determined fiducial IMF for our model. For comparison we also show the log-likelihood of the observational data with synthesised data generated from our model but using common IMFs from the literature.
\subsection{New IMF parameters}
\label{sec:our_IMF}
 \begin{figure}
 \includegraphics[width=0.495\textwidth]{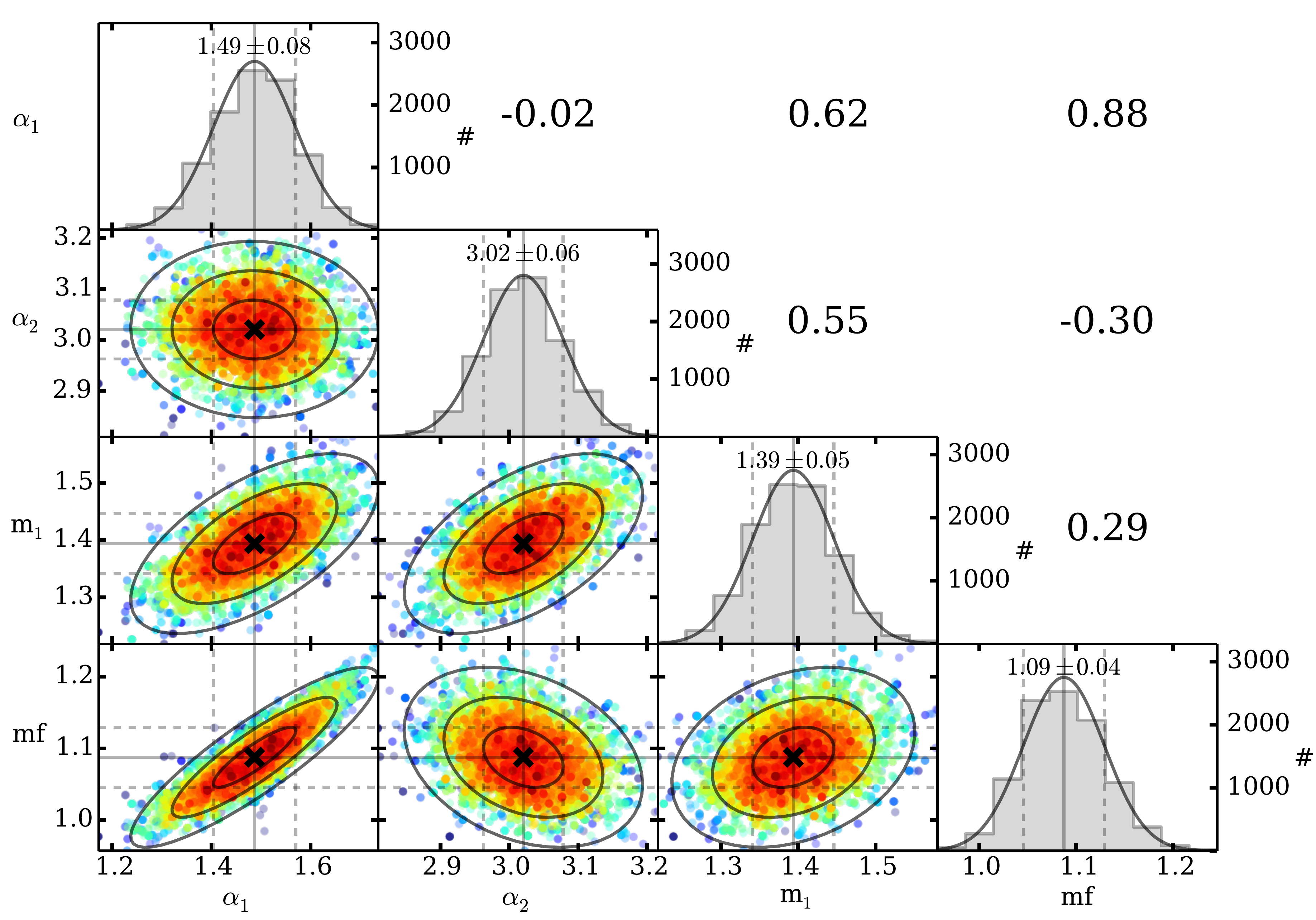}
 \caption[parameter space]{Marginalised parameter distribution of the MCMC run exploring the equilibrium distribution of the parameter space with respect to our log-likelihood using $10^5$ evaluations. Each scatter plot shows the projected 2d parameter distribution with points coloured by likelihood increasing from blue to red. Crosses indicate the mean values and ellipses encompass the 1-3\,$\sigma$ regions. The respective correlation coefficients are given at the position mirrored along the diagonal. Gaussian fits and histograms of the marginalised parameter distribution are given on the diagonal. The mean and standard deviation of each parameter is written and also indicated by solid and dashed grey lines.}
 \label{fig:parameter}
 \end{figure}
  \begin{figure}
 \includegraphics[width=0.48\textwidth]{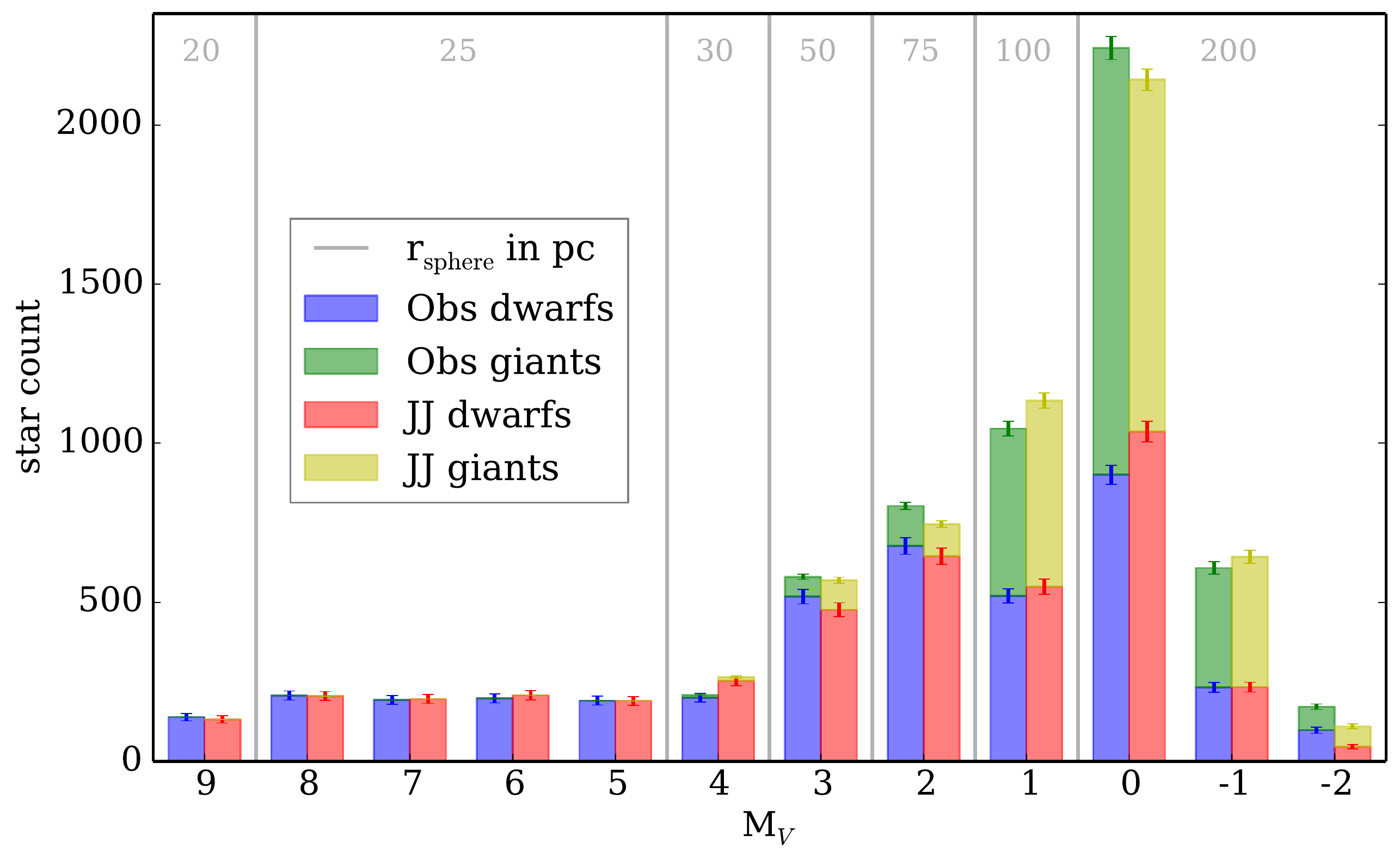}
 \caption[Luminosity function]{Luminosity function of the observations and the new IMF of the JJ model (cf. table\,\ref{tab:hipparcos_dwarf} and\,\ref{tab:hipparcos_giant}). Error bars indicate Poisson noise in the observational sample and the standard deviation for the 400\,times oversampled synthesised catalogue. Star counts are not normalised for the different distance limits. The limiting radii of the corresponding magnitude bins are written in the top.}
 \label{fig:lf}
 \end{figure}
Multiple burn-ins from different starting points all settling in the same equilibrium configuration suggest a well-behaved parameter space with respect to our likelihood calculation. We use 20\,walkers each sampling 500\,steps to generate the point cloud representing the PDF of the parameter space. For each step we average over 400\,realisations. Figure\,\ref{fig:parameter} shows the marginalised likelihood distribution of each parameter as histograms on the diagonal and as point clouds for each parameter pair in the lower left. Just for illustrative reasons each dot is coloured to indicate high ($\mathrm{red}_\mathrm{max}=-77.2$) and low ($\mathrm{blue}_\mathrm{min}=-86.7$) log-likelihood values with grey dots being below this 3\,$\sigma$ range. In the histograms the mean and the standard deviation of the number density of each marginalised parameter is given which represents our central result and defines our new fiducial IMF: 
\begin{equation}
\begin{array}{rcl}
\alpha_1 &=& 1.49\pm 0.08,\\
\alpha_2 &=& 3.02\pm 0.06,\\
\mathrm{m}_1 &=& 1.39\pm 0.05,\\
\mathrm{mf} &=& 1.09\pm 0.04.\\
\end{array}
\end{equation} 
Furthermore Pearson's correlation coefficient for each parameter pair is given in the upper right of figure\,\ref{fig:parameter} which is also represented in the 1-3\,$\sigma$\,ellipses of the projected point clouds. The only two parameter which are uncorrelated are the power law indices. Almost positive linear is the correlation of the mass factor with the low mass slope. This is due to more mass being put into stars which are not represented in our data ($m<0.5$\,M$_\odot$) for high $\alpha_1$ which can be counterbalanced with a high mass factor. Similarly but less strong the anti-correlation of the mass factor with the high mass slope is due to mass being shifted out of our represented data domain when $\alpha_2$ is getting lower. The reason for both power law indices being positively correlated with the power law break is due to the shape of the IMF which is sharply decreasing at the position of the break and the number densities of the faintest and brightest magnitude bin which need to be matched when shifting $\mathrm{m}_1$. This reasoning can be best visualised when looking at the green dotted line (our IMF) and the blue error bars (our observational sample) in figure\,\ref{fig:lf}. Additionally when increasing both $\alpha_1$ and $\alpha_2$ the mass that is gained by the steeper low mass slope will be counterbalanced by the mass loss of the steeper high mass slope.\\ 
Figure\,\ref{fig:lf} shows the binned luminosity function of our fiducial IMF (averaged over 400\,realisations) compared to the observations in absolute numbers. The CMD representation of the luminosity functions can be seen in figure\,\ref{fig:division} where the same colours have been used for the different samples. The deviations in each bin look small and systematics are not apparent neither in the dwarf nor the giant sample which shows that the whole machinery, consisting of the disc model and Galaxia producing the mock observations, works well and the MCMC simulation has likely converged towards the equilibrium configuration.\\
When inspecting the detailed likelihood contribution of each bin in table\,\ref{tab:hipparcos_dwarf}\,\&\,\ref{tab:hipparcos_giant} we see that the largest penalty comes from the 0th VMag bin. Especially the giants with $\ln\left(\mathcal{L}/\mathrm{P}_\mathrm{max}\right)=-23.4$ have a huge impact. The reason for that is likely a red clump that is too faint in our mock catalogue. In the dwarf sample too many stars are in the synthesised 0th VMag bin with JJ$_{25}\cdot$(N$_{fin}$/N$_{25})=1036$ compared to 901\,stars in the Hipparcos catalogue.\\
A weakness of our modelling machinery is apparent in the high $\ln\left(\mathcal{L}/\mathrm{P}_\mathrm{max}\right)$ value of the brightest dwarf bin indicating that too few bright stars are produced. The reason is most likely that we are not accounting for binaries. Minor effects could be missing high metallicity stars and that our synthesised stars are not younger than 6.25\,Myr. It could also indicate a change of the high mass power law index for stars more massive than contained in our data.
\subsection{Tested IMFs}
\label{sec:otherIMF}
Because for other shapes of the IMF part of the mass could also be hidden in the mass range not represented by our observational sample (0.5 to 10\,M$_\odot$) we determine the factor with which the total mass needs to be rescaled in order to maximise the likelihood for a particular IMF when it is used with our JJ-model against the observational sample. We again average over 400\,realisations and get the standard deviation as the enclosing 68\,\% of the likelihood.\\
In table\,\ref{tab:IMFcomp} the log-likelihoods of the different IMFs (using our disc model but adjusting for each IMF's best fit mass factor) are listed with JJ$_{3\sigma}$ being the log-likelihood value which is lower than 99.7\,\% of the points representing the PDF in figure\,\ref{fig:parameter}.\\
Table\,\ref{tab:bootstrap} illustrates a few properties of our log-likelihood. JJ$_{400}$ and JJ$_{1}$ show the mean and the standard deviation of 100 log-likelihood determinations with different seeds which are averaged over 400 in the former and 1\,realisation in the latter case. This shows that the averaging is important for the MCMC simulation in order to smoothen the likelihood distribution. The deteriorated mean for single realisations is due to a skewed distribution since P$_\mathrm{max}$ is a lower limit and the increasing penalty for extreme values.\\
The last column of table\,\ref{tab:bootstrap} (JJ$_\mathrm{ideal}$) gives an ideal log-likelihood which is obtained when the data is indeed represented by the model. For that we draw 100 single random samples (JJ$_1$) from JJ$_{400}$ with replacement and evaluate their log-likelihood with the parent distribution (JJ$_{400}$). Each sample fulfilling the observational constraint of having dwarf and giant star counts fixed to N$_\mathrm{Obs,dwarf}=4075$ and N$_\mathrm{Obs,giant}=2513$. This means that a $\ln\left(\mathcal{L}/\mathrm{P}_\mathrm{max}\right)$ of around $-13.8$ would indicate a perfect model. An even lower log-likelihood value close to $\mathrm{P}_\mathrm{max}$ ($\ln\left(\mathcal{L}/\mathrm{P}_\mathrm{max}\right)=0$) would be unrealistic since there is a natural scatter to Poisson processes.\\
Related to that we also inspected the distribution of star counts in individual magnitude bins for random realisations (with the same parameters but different seeds) which indeed is Poissonian.
\begin{table}
\caption{Likelihoods of the different IMFs}
\begin{center}
\begin{tabular}{ c|c c c c c}
  IMF &  JJ$_\mathrm{}$ & JJ$_{3\sigma}$  & Bes\,B & KTG\,93 & Cha\,03  \\
\hline  
  $\ln\left(\frac{\mathcal{L}}{\mathrm{P}_\mathrm{max}}\right)$ & $-79.3$ & $-86.7$ & $-96.7$ & $-195.6$ & $-216.1$  \\
\end{tabular}
\end{center}
\label{tab:IMFcomp}
\end{table}
\begin{table}
\caption{Variability of the log-likelihood}
\begin{center}
\begin{tabular}{ c|c c c}
    Sample &  JJ$_{400}$ & JJ$_{1}$  & JJ$_\mathrm{ideal^\star}$  \\
\hline  
  $\ln\left(\frac{\mathcal{L}}{\mathrm{P}_\mathrm{max}}\right)$ & $-79.3\pm0.6$ & $-89.5\pm14$  & $-13.8\pm2.8$ \\
\end{tabular}
\begin{tablenotes}
\item$^\star$if data was coming from our model (see text)
\end{tablenotes}
\end{center}
\label{tab:bootstrap}
\end{table}
\subsubsection*{KTG\,93}
The widely used \citet{Kr93} IMF is a three slope broken power law of the form 
\begin{equation}
\frac{\mathrm{d}n}{\mathrm{d}m}=k_\alpha m^{-\alpha}
\left\{ \begin{array}{l}
 \alpha = \alpha_1, \\
 \alpha = \alpha_2, \\
 \alpha = \alpha_3,
 \end{array} \mbox{\ if \ } 
 \begin{array}{l}
 \mathrm{m}_\mathrm{low}<m<\mathrm{m}_1\\
 \mathrm{m}_1<m<\mathrm{m}_2\\
 \mathrm{m}_2<m<\mathrm{m}_\mathrm{up}\end{array} \right.
\label{KTG}
\end{equation}
with $\alpha_1=1.3,\alpha_2=2.2, \alpha_3=2.7,\mathrm{m}_1=0.5$ and $\mathrm{m}_2=1$. Again $k_\alpha$ is a normalisation constant ensuring a continuous IMF between $\mathrm{m}_\mathrm{low}=0.08$\,$\mathrm{M}_\odot$\,and\,$\mathrm{m}_\mathrm{up}=100\,\mathrm{M}_\odot$.\\
The likelihood peak is obtained with a mass factor of $1.392\pm0.019$ which is quite high. The reason for that is too much mass being put into low mass stars which are not represented in our observational sample (see table\,\ref{tab:sn2}).\\
With $\ln\left(\mathcal{L}/\mathrm{P}_\mathrm{max}\right)=-195.6$ it scores poorly compared to our or the Besan\c{c}on\,B IMF showing that its shape is not able to reproduce local star counts. This is visible in figure\,\ref{fig:imf} where it produces too few stars within 0.9 - 2\,M$_\odot$ and too many outside of this range compared to our IMF.
\subsubsection*{Besan\c{c}on\,B}
This is one of the new fiducial Besan\c{c}on model IMFs from \citet{Cz14} tested with Tycho\,2 all-sky colour distribution. It is also a three slope broken power law with $\alpha_1=1.3,\alpha_2=1.8, \alpha_3=3.2,\mathrm{m}_1=0.5$ and $\mathrm{m}_2=1.53$.\\
The mass factor for this IMF is $1.107\pm0.015$ which is compatible with our own $\mathrm{mf}$. Also the shape of the IMF (cf. figure\,\ref{fig:imf}), the likelihood (see table\,\ref{tab:IMFcomp}) and the mass fractions (cf. table\,\ref{tab:sn2}) are similar. This is not surprising since they are also using the local mass density model from \citet{Ja97}. Still it is reassuring that for different observational data (Tycho\,2 colour vs. Hipparcos/CNS5 VMag) and different modelling techniques similar results are yielded. Although the difference of 17.4 in $\ln\left(\mathcal{L}/\mathrm{P}_\mathrm{max}\right)$ still means that the likelihood for their IMF is 36\,million times lower than for our IMF.\\
Compared to our IMF the Besan\c{c}on\,B IMF is producing slightly less high mass stars and more low mass stars which could be partly due to their rigorous treatment of binaries (see figure\,\ref{fig:imf} and cf. section\,\ref{sec:binary}).
\subsubsection*{Chabrier\,03}
Another widely used IMF comes from \citet{Ch03}. It is a mixture of a lognormal form in the low mass and a Salpeter power law in the high mass regime
 \begin{equation}
 \frac{{\mathrm d}n}{{\mathrm d}m}=\left\{ \begin{array}{l}
 \frac{0.852464}{m}\,\exp\big(\frac{-\log^2\left(\frac{m}{0.079}\right)}{2\cdot 0.69^2}\big), \\
 0.237912\cdot m^{-2.3},
 \end{array} \mbox{\ if \ } \begin{array}{l}m<\mathrm{M}_\odot\\ m>\mathrm{M}_\odot .\end{array} \right.
\label{eq:Cha} 
 \end{equation} 
The mass factor for the best likelihood is: $1.317\pm0.017$. It is so high because a huge mass fraction is going into stars more massive than 8\,M$_\odot$ (cf. table\,\ref{tab:sn2}, highest supernova rate compared to any other standard IMF). The shape of the Chabrier\,03 IMF fits our model worst with respect to the data scoring a log-likelihood of $\ln\left(\mathcal{L}/\mathrm{P}_\mathrm{max}\right)=-216.1$.\\
\subsection{From luminosity function to local stellar mass density}
\label{sec:LF}
  \begin{figure}
 \includegraphics[width=0.48\textwidth]{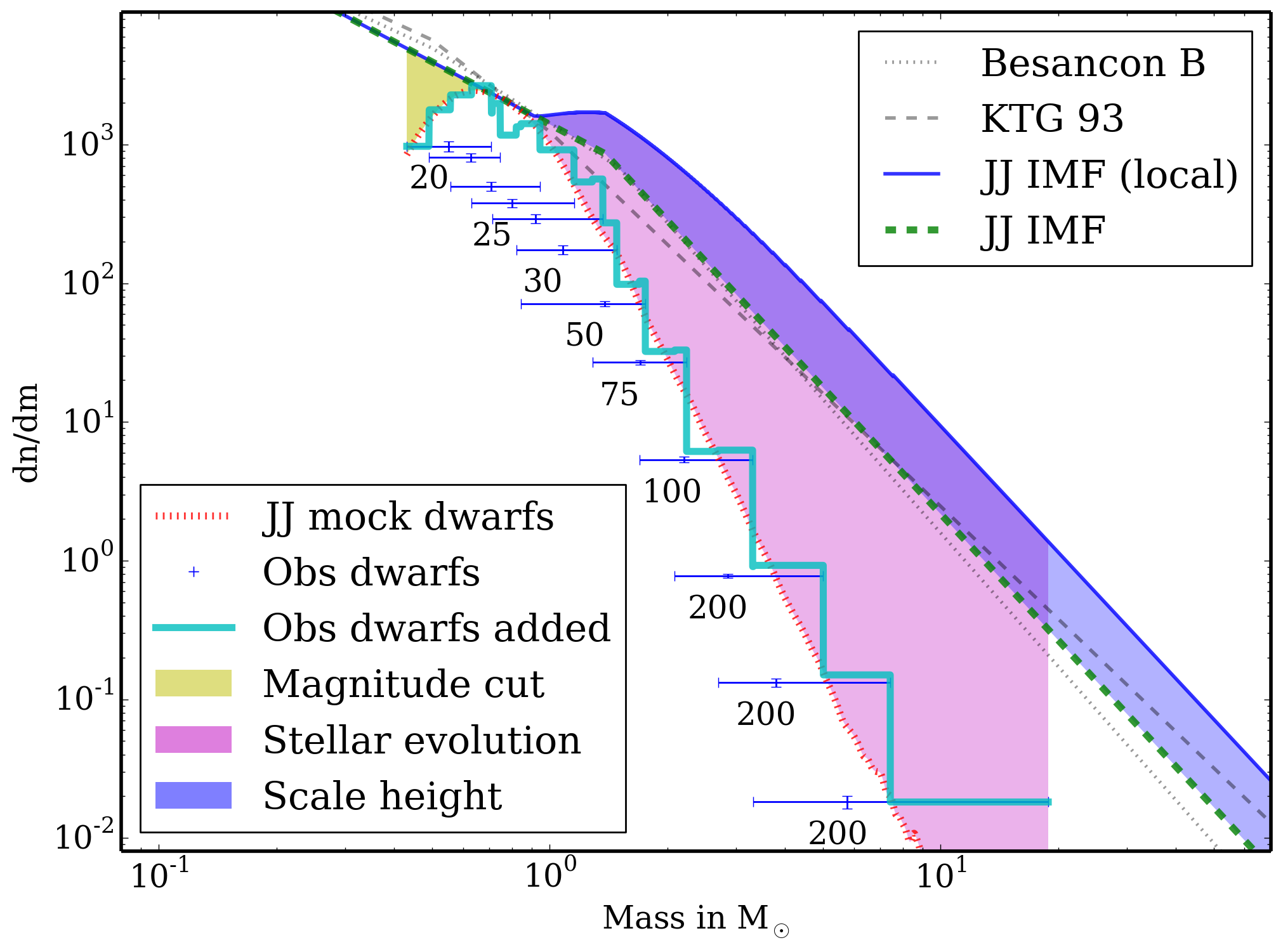}
 \caption[IMF]{25\,pc LF of dwarf stars translated into mass space. Our fiducial IMF is plotted in thick dashed green and the Besan\c{c}on\,B and KTG\,93 IMFs are plotted in dotted and dashed grey. The blue error bars represent the 12\,magnitude bins from the observational dwarf sample with their limiting radius given in parsec. The x-value and the x-error are associated with the median and the range of stellar masses in the corresponding mock magnitude bin synthesised with our fiducial IMF. The y-value represents the number of stars normalised to the mass range and the y-error is the Poisson noise. As the masses from different magnitude bins overlap the values are added up in the solid cyan line representing a kind of {\it observational} PDMF. Below 0.7\,$\mathrm{M}_\odot$ the yellow shaded area indicates the mass that is missing due to the magnitude cut. In dotted red the same effect is visible as this line represents the synthesised mass function of our IMF averaged over 100\,realisations. From 0.9\,M$_\odot$ upwards the IMF and the local IMF in solid blue deviate since locally (i.e. close to the Galactic plane) young and therefore massive stars are over represented which is indicated by the blue shaded area. Therefore the PDMF in the 25\,pc sphere would look like the local IMF if there was no stellar evolution (shown in magenta) at work.}
 \label{fig:imf}
 \end{figure}
In figure\,\ref{fig:imf} the IMFs are normalised such that their integrated mass is representing the mass of gas that was turned into stars (of which a few already turned into remnants) still residing in the 25\,pc sphere (i.e. $\mathcal{M}_\mathrm{IMF,25\,pc}$ including the thick disc and each IMF's mass factor cf. equation\,\ref{eq:totalmass}). Beware that in figure\,\ref{fig:imf} actually the number of stars per mass interval is displayed though normalisation happens in mass space.\\
The yellow shaded area shows the incompleteness of our observations in low mass stars due to the magnitude cut. This comes from high metallicity stars being redder and fainter than their equal mass low metallicity counterparts so they are excluded and do not contribute to the smallest mass bin leading to a decrease of the observed mass compared to the theoretical IMF. The same is true for the JJ mock dwarfs in dotted red (as we apply the same magnitude cuts) which were spawned using the green dashed IMF. We could drop the magnitude cut for faint stars in Galaxia and the red dotted line would be perfectly aligned with the JJ IMF in the low mass regime which corresponds to the local present day mass function (PDMF). In the {\it observational} PDMF represented by the solid cyan line the little bump compared to the JJ mock dwarfs at around 0.8\,M$_\odot$ could be related to the isochrones badly fitting our low mass stars biasing the associated mass ranges.\\
In the high mass regime the blue shaded area shows the over representation of massive stars in the local IMF of the 25\,pc sphere as their vertical distribution is more confined to the Galactic plane (if one would not account for the scale height dilution the deduced IMF would look like this). The magenta area indicates the deviation from the local IMF to the mock sample due to stellar evolution.\\
We can quantify the difference of IMF and PDMF by looking at the integrated mass of the two functions represented by \textit{JJ IMF} and \textit{JJ mock dwarfs} from figure\,\ref{fig:imf} (the latter being made equal to \textit{JJ IMF} for the low mass part not affected by stellar evolution). The outcome is that 55.8\,\% of the mass originally turned into stars is still present in dwarf stars today ($\mathcal{M}_{\mathrm{PDMF,25\,pc}}/\mathcal{M}_{\mathrm{IMF,25\,pc}}$ ). Including giants this value only changes slightly to 56.9\,\% which is consistent with the integrated stellar mass of our disc model (figure\,1 of paper\,I adapted for 25\,pc age distribution yields 56.5\,\% of stars increasing to 70.8\,\% when remnants are included which is their g$_\mathrm{eff}$) confirming that the new IMF still fits within our model's framework as SFH and AVR are only dependent on the integrated mass loss. 

In order to clarify our remnants which Galaxia is not synthesising we derive from our PDMF and IMF that we should have 707 stellar remnants for stars between 1 and 8\,M$_\odot$ (potentially WDs) and 17 heavier ones (potentially black holes or neutron stars) in the 25\,pc sphere. Compared to \citet{Si14} we have over a factor of 2 more since they expect 344\,WDs within the same limits. But their assumed volume completeness for the 13\,pc sample seems quite optimistic at least when speaking of the cool end of the WD cooling sequence. \citet{Ho08} propose a WD mean mass of 0.665\,M$_\odot$ which for us result in a WD mass density of 0.0072\,$\mathrm{M}_\odot/\mathrm{pc}^3$ quite close to the Besan\c{c}on value based on \citet{Wi74} which was already corrected downwards in \citet{Ja97} because one out of 5\,WD left the 5\,pc sphere. We admit that our high number of WDs is partly due to the two slope IMF structure having a bump where the power law break lies (when expecting a concave function describing the underlying distribution in log-space) which slightly exaggerates the mass fraction of the IMF going into planetary nebulae (cf. table\,\ref{tab:sn2}).

With these values (summarised in table\,\ref{tab:Holmberg}) the overall present day mass fraction of stars and stellar remnants ($\mathcal{M}_\mathrm{PDMF,total,\,25\,pc}$) is
\begin{equation}
g_\mathrm{eff} = \frac{\mathcal{M}_\mathrm{PDMF,\,25\,\mathrm{pc}}+\mathcal{M}_\mathrm{BD\,\& WD\,25\,\mathrm{pc}}}{\mathcal{M}_\mathrm{IMF,25\,pc}} = 71.7\,\%
\end{equation}
which is close to 70.8\,\% for the 25\,pc sample of paper\,I. This also implies a combined BD and WD mass fraction of about $20\,\%$ of the local mass budget ($\mathcal{M}_\mathrm{PDMF,total,\,25\,pc}$) consistent with the \citet[tab.\,3]{Ja97} value and the stellar evolution of paper\,I. Then again we could also have chosen the proposed WD local mass density of \citet{Ho08} $\mathrm{M}_\odot/\mathrm{pc}^3 = 0.0032$ which would decrease the local disc mass density and change g$_\mathrm{eff}$ and the remnant fraction.\\
\begin{table}
\caption{Local stellar mass density of different thin disc components in $10^{-4}\mathrm{M}_\odot/\mathrm{pc}^3$}
\begin{center}
\begin{tabular}{ c|c c}
Mass component  & \citet{Fl06} & JJ-model \\
\hline
giants  & 6   & 7  \\
\phantom{$2..5<$}$\mathrm{M}_\mathrm{V}<2.5$  & 31  & 11 \\
$2.5<\mathrm{M}_\mathrm{V}<3$\phantom{$.5$} & 15  & 5  \\
$3<\mathrm{M}_\mathrm{V}<4$   & 20  & 19 \\
$4<\mathrm{M}_\mathrm{V}<5$   & 22  & 26 \\
$5<\mathrm{M}_\mathrm{V}<8$ & 70  & 66 \\
$8<\mathrm{M}_\mathrm{V}$\phantom{$5.<$}  & 135   & 205\\
\hline

$\rho_\mathrm{thin\,disc}(z=0\,\mathrm{pc},t=12\,\mathrm{Gyr})$ & 299 & 338\\
\hline
$\rho_\mathrm{thick\,disc}(z=0\,\mathrm{pc},t=12\,\mathrm{Gyr})$ & 35 & 17\\
\hline
brown dwarfs    & 20  & \phantom{$^*$}20$^\star$ \\
white dwarfs    & 60  & \phantom{$^*$}72$^\dagger$ \\
\hline
\hline
$\rho_{\mathrm{PDMF,\,total}}$ & 414 & 447 \\
\end{tabular}
\begin{tablenotes}
\item $^\star$taking the same value as FH\,06 
\item $^\dagger$derived implicitly from our PDMF (see text)
\end{tablenotes}
\end{center}
\label{tab:Holmberg}
\end{table}
For the following local mass density test we use our disc model and the fiducial IMF to produce all stars within 25\,pc down to 0.08\,M$_\odot$ without any magnitude cuts and analyse the sample's properties. In this sample we find a local stellar mass density of 0.034\,$\mathrm{M}_\odot/\mathrm{pc}^3$ compared to 0.030\,$\mathrm{M}_\odot/\mathrm{pc}^3$ for the same selection of stars in \citet[tab.\,2]{Fl06} (including giants, excluding BDs, WDs, other remnants and the thick disc component) who use a similar method. In table\,\ref{tab:Holmberg} the detailed comparison reveals that especially the two brightest bins of the FH\,06 sample are 3\,times denser than the corresponding bins in the JJ sample. For $\mathrm{VMag}>3$ the densities are matched quite well except for the faintest mag bin. The over abundance of bright stars could be coming from their large value of n in \citet[eq.\,3]{Ho97} resulting in nearly exponential vertical density distribution with high local mass densities though this then should apply to the other mag bins as well. Another indication for an over estimation of their model's local star counts becomes evident when comparing the number densities from \citet{Ho00} (upon which FH\,06 is based) with the CNS\,5 ones. For $\mathrm{M}_\mathrm{V}<2.5$ they have 0.0013\,star\,pc$^{-3}$ whereas our volume complete sample has half of it with 0.0007\,star\,pc$^{-3}$ and this despite the fact that the 25\,pc from the CNS\,5 seem to have an over representation of the upper MS compared to the larger sample (cf. table\,\ref{tab:hipparcos_dwarf}). The next mag bin $2.5<\mathrm{M}_\mathrm{V}<3$ is three times denser according to \citet{Ho00} with 0.0010\,star\,pc$^{-3}$ compared to 0.0003\,star\,pc$^{-3}$ we measure for our 25\,pc sample which strongly indicates a necessary revision. Fainter mag bins are much better fit in star counts as well as in stellar mass density.\\
Comparing our thin disc stellar mass density (excluding thick disc stars and WDs) of 0.0338\,$\mathrm{M}_\odot/\mathrm{pc}^3$ to the default Besan\c{c}on\,B model that has 0.0330\,$\mathrm{M}_\odot/\mathrm{pc}^3$ reveals a similar discrepancy. Our disc model has 12\,Gyr of evolution compared to 10\,Gyr and we make a detailed comparison in table\,\ref{tab:mass_density} keeping in mind the different SFH and vertical profiles of each disc model \citep[tab.\,7 and fig. 4]{Cz14}.
The Besan\c{c}on\,B model use the same local mass density as we do from \citet{Ja97} based on 25\,pc though they add the thick disc on top of this value which is an inconsistency as the \citet{Ja97} table\,3 accounts for all local stars not only the thin disc.\\
\begin{table}
\caption{Local mass density over stellar age in $10^{-4}\mathrm{M}_\odot/\mathrm{pc}^3$}
\begin{center}
\begin{tabular}{ c|c c c}
Age [Gyr] & Besan\c{c}on\,A & Besan\c{c}on\,B & JJ-model \\
\hline  
\phantom{0.15}0 - 0.15\phantom{0} & 20 & 19 & 9 \\ 
\phantom{1}0.15 - 1\phantom{0.15} & 55 & 50 & 36 \\ 
1 - 2 & 46 & 41 & 30 \\
2 - 3 & 33 & 28 & 27 \\
3 - 5 & 58 & 49 & 52 \\
5 - 7 & 61 & 50 & 54 \\ 
\phantom{10}7 - 10\phantom{7} & 117 & 93 & 84 \\
\phantom{12}10 - 12\phantom{10} & - & - & 46 \\
\hline
Thin disc & 390 & 330 & 338 \\
Thick disc &29&29&17\\
\hline
WD & 71 & 71 & \phantom{$^\dagger$}92$^\dagger$ \\
\hline\hline
$\sum$ & 490 & 430 & 447 \\
\end{tabular}
\begin{tablenotes}
\item $^\dagger$ including brown dwarfs
\end{tablenotes}
\end{center}
\label{tab:mass_density}
\end{table}
Overall for the other disc mass models more mass seems to be sitting in stellar mass bins of bright stars which might be partly due to the local dwarf sample being almost 2\,times denser than the 200 and 100\,pc sample at the upper MS (cf. table\,\ref{tab:hipparcos_dwarf} {\it Observations} \& {\it CNS\,5}). Since we use volume complete samples to fit the luminosity function and take into account the scale height dilution according to the stellar ages we trust our mass distribution in the mass range that our observational evidence samples ($0.5-10$\,M$_\odot$). Ideally we should have included a constraint for the low mass stellar mass density from other observations because now the mass factor and the low mass power law index are strongly correlated and certainly a bit too high. Better values for a more realistic two slope IMF would probably be in the lower left of the 1\,$\sigma$ ellipse in figure\,\ref{fig:parameter} (e.g. $\mathrm{mf}\simeq1.05$ and $\alpha_1\simeq1.4$).\\ 
\section{Discussion}
\label{sec:disc}
We discuss our findings with respect to their model dependencies as well as our analysis method. Then we provide a comparison of the mass distribution from various IMFs and end this section with an empirically driven adaptation of our IMF to account for missing low mass star representation in our data.
\subsection{Mass factor}
The overall mass with our newly determined IMF compared to the JJ-model SFH increased by the thick disc fraction (6.5\,\%) and the mass factor ($\mathrm{mf}=1.09$). The normalisation in paper\,I was done using $\rho_\mathrm{PDMF,total}=0.039\,\mathrm{M}_\odot$pc$^{-3}$ from \citet{Ja97}. Since we utilise new isochrones together with number densities derived from volume complete star counts a change of about 10\,\% is not unexpected but our value of $0.045\,\mathrm{M}_\odot$pc$^{-3}$ is probably exaggerated. The present day mass fraction (g$_\mathrm{eff}$) stays similar and also the remnant fraction is compatible with paper\,I values as shown in section\,\ref{sec:LF} though WD and faint stellar local mass density are arguably too high. The problem is that we do not have observational constraints for the whole mass range resulting in a degeneracy of the mass factor and the low mass slope (see figure\,\ref{fig:parameter}, also valid for the high mass slope). We propose a solution to this in section\,\ref{sec:empirical_fix}.
\subsection{Isochrones}
\label{sec:isochrones}
A crucial ingredient for our investigation is the used set of isochrones since it translates our analytical disc model into the realm of observables. When we were using the option provided originally by Galaxia \citep{Ma08} the highest likelihood we could score was $\ln\left(\mathcal{L}/\mathrm{P}_\mathrm{max}\right)=-109$ with slightly different IMF parameters. With the latest PARSEC isochrones the observations are much better fit by the model increasing the normed log-likelihood by 30. Still a few discrepancies are visible when inspecting figure\,\ref{fig:division}.\\
The cold dwarf V-Band problem was already mentioned in section\,\ref{galaxia} and is discussed in \citet{Ch14}. We would not recommend to go much fainter in V band for such an analysis.\\
Section\,\ref{sec:our_IMF} discusses the huge likelihood penalty from the brightest magnitude bin in table\,\ref{tab:hipparcos_dwarf} which could be due to unaccounted binaries in our sample, missing turn-off stars (which could be related to our discrete time steps), missing super solar metallicity stars or a high mass power law index which is too large.\\
Table\,\ref{tab:hipparcos_giant} shows missing stars in the 0th mag bin and an over abundance in the 1st for the giants. This is an indication for the red clump being too faint compared to the observations.\\
Another striking feature is the complete absence of synthesised stars blueward from the red clump whereas Hipparcos shows several dozens. Maybe variable stars are not well represented in our isochrones or it could again be a discrete time sampling effect.\\
Of course not all differences are linked to the isochrones as our model SFH, AMR or IMF also affect the distribution of stars in the CMD. Another reason for mismatch is our used reddening law not accounting for inhomogeneous ISM which is probably the reason for all the unmatched faint giants in figure\,\ref{fig:division}.\subsection{Binarity}
\label{sec:binary}
In our observational sample we tried to account for all binaries that have $\Delta \mathrm{mag}$\,entries so that we could split them up. Apart from these another 365\,stellar entries in our observational sample are listed in the \textit{Washington Double Star Catalogue} \citep{Ma01} or the \textit{Catalogue of Components of Double and Multiple Stars} \citep{Do02} as multiple systems but are not resolved (i.e. they are not split up for this analysis but just kept as one {\it star}).\\
As mentioned in section\,\ref{sec:our_IMF} we have a problem with binaries in massive stars which are hard to detect as lines are blurred and increased luminosity could also be due to ageing of a single star. Since binary fraction in massive stars is high and our likelihood gets a large penalty from the brightest dwarf mag bin not being matched well we believe that we miss quite a few high mass binaries. Other than that listings of binary stars in the Hipparcos catalogue \citep{Pe97} are said to be {\it fairly complete} \citep{Li97} for $\Delta Hp<3.5$ and an angular separation bigger than 0.12 to 0.3\,arcsec (increasing with $\Delta Hp$). 
In CNS\,5 all stellar systems from the literature with resolved magnitude differences are split into their respective components.\\
The Besan\c{c}on group is doing the favourable approach of accounting for binaries from the models side \citep{Cz14,Ro12,Ar11}. They use an angular resolution limit of 0.8\,arcsec for resolved binaries in their Tycho\,2 data. In \citet[fig. A.3]{Cz14} the relative difference in stellar mass frequency produced with binarity treatment in single, primary and secondary stars compared to the same sample excluding secondary stars is shown. With B-components included the IMF produces around 6\,\% more stars below 1.1\,M$_\odot$ and around 6\,\% less above with a short transition in between. Overall the effect of binarity seems to play a secondary but not negligible role especially in massive dwarf stars.\\
\subsection{Splitting the CMD}
Before exploiting the full 2d information of the CMD and adapting the statistical machinery as well as dealing with colour issues of the isochrones (not speaking of enhanced sensitivity to reddening) the easiest way to increase the data constraints is to split up the CMD into dwarf and giant sample. Weighting both into the final likelihood is a valuable gain since they represent stellar populations with different ages and masses (check the last columns of table\,\ref{tab:hipparcos_dwarf}\,\&\,\ref{tab:hipparcos_giant}) and are still build from the same IMF (as well as SFH, AVR and AMR). Interestingly the likelihood penalty from both samples are similar though the dwarfs contribute 12 and the giants only 7\,bins. This probably shows the challenge stars on the giant branch still pose to stellar evolutionary modelling.\\
Exemplary for the insight gained from splitting up the CMD the 0th mag bin can be inspected where a common sample would have balanced our model predicting too few giants and too many dwarfs. As these two bins have well distinct ages and masses constraints are put on different parts of the IMF. On the other hand the over abundance of stars in the 3rd mag bin in giants and the 4th mag bin in both giants and dwarfs is indicating too many stars for the IMF around M$_\odot$. This is in balance with the depletion of stars from around 1.3\,M$_\odot$ (2nd mag in giants and 3rd in dwarfs) in order to fit the two slope power law with respect to our constructed likelihood.
\subsection{High mass slope}
\label{sec:highmass}
Our high mass power law index of $\alpha=3.02$ is at the higher end of the literature values which can be partly attributed to the power law break being at a comparatively high mass. Still recent studies converge towards a steeper high mass slope as for example both new default IMFs of the Besan\c{c}on model \citep{Cz14} have similar values.\\
An important measure for an IMF in terms of Galaxy simulation is the fractional mass going into stars heavier than 8\,M$_\odot$ as they supposedly explode as supernova of type\,II (SN\,II) and have the highest and fastest stellar feedback to the Galactic evolution in terms of gross elemental synthesis and heating of the gas phase. In table\,\ref{tab:sn2} we list the fraction of total mass and total star count going into the SN\,II mass bin for all investigated IMFs and Salpeter with $\alpha=2.3$:
\begin{table}
\begin{center}
\caption{Mass distribution, planetary nebula and supernova type\,II occurrences for different IMFs}
\begin{tabular}{ c|c c c c c }
IMF:    &JJ   & Bes\,B& KTG\,93& Cha\,03& Sa\,55\\
\hline
Mass range in M$_\odot$  & \multicolumn{5}{|c|}{mass fraction in\,\%} \\   
\hline  
\phantom{100}$8-100$\phantom{8}   & 6   & 4   & 8   & 22  & 15  \\
\phantom{8}$1.4-8$\phantom{1.4}  & 32  & 30  & 24  & 29  & 20  \\
\phantom{1.4}$1-1.4$\phantom{1}  & 13  & 12  & 9  & 7   & 5   \\
\phantom{1}$0.5-1$\phantom{0.5}   & 20  & 23  & 22  & 16  & 12  \\
\phantom{0.5}$0.08-0.5$\phantom{0.08}  & 29  & 31  & 36  & 26  & 48  \\
\hline
\hline
PN$^\star$ ($1-8$)   & 994   & 951   & 707   & 687   & 467   \\
SN\,II$^\star$ ($>8$)  & 17  & 12  & 21  & 47  & 32  \\
\end{tabular}
\begin{tablenotes}
\item $^\star$number of occurrences for an SSP with mass M$_\mathrm{IMF,25\,pc}\approx4100\,\mathrm{M}_\odot$ at $\mathrm{t}=\infty$.
\end{tablenotes}
\label{tab:sn2}
\end{center}
\end{table}
The fraction of mass going into stars ending their lives as SN\,II ranges from 4\,\% for Besan\c{c}on\,B \citep{Cz14} to 22\,\% for \citet{Ch03}.\\
In order to obtain sensible results modellers usually tune their feedback physics according to their used IMF. Chemical enrichment models (CEMs) and their predicted element ratios are especially sensitive to the number of SN\,II events and the high mass slope. So using a Chabrier IMF will need substantially different stellar yields to obtain similar results (i.e. reproduce observations) compared to an CEM using the new Besan\c{c}on\,B IMF. Same is true for cosmological hydrodynamical simulations trying to reproduce the number of dwarf galaxies around spirals and so forth.\\
Still our high mass slope is an extrapolation for stars heavier than about 10\,M$_\odot$ as massive stars are too rare and not homogeneously distributed enough to be represented by our observational sample limited to 200\,pc. The deficit of stars in our brightest dwarf bin might not only be due to missing binaries but also indicate a higher abundance of high mass stars than represented by our two slope IMF. 
\subsection{Empirically motivated 3 slope IMF extension}
\label{sec:empirical_fix}
Our observational sample not constraining the high (volume too small) and low mass stars (sparse V Band data and no reliable isochrones) is deteriorating our inferred IMF parameters. A good solution would be to include observationally based priors into the likelihood determination directing the MCMC simulation to a more physically motivated solution. Possible constraints directly connected to the shape of the IMF and the mass normalisation could be the SN\,II rate, H\,II\,regions, PN rate, WD number density, low mass stellar density. But it is not trivial to account for the uncertainty of those observations in the prior function.\\
As an empirical fix we introduce a second power law break at 0.5\,M$_\odot$ leaving the shape of the IMF above the same but changing it for lower masses. With a look at table\,\ref{tab:Holmberg} all bins should stay the same except for the $8<\mathrm{M}_\mathrm{V}$ bin which should decrease to $0.017\,\mathrm{M}_\odot/\mathrm{pc}^3$ being a new value derived from volume complete near infrared data of the CNS\,5 which will soon be published (Just et al. 2015, in preparation). This deflates our mass model but still uses our high quality data for the other mass bins. The proposed IMF parameters fulfilling this additional constraint are: $\alpha_0=1.26$, $\mathrm{m}_0=0.5$, and $\mathrm{mf}=1.03$. Compared to our low mass slope of $\alpha_1=1.49$ the extension to lower masses is a bit shallower and quite similar to KTG\,93 also having its low mass power law break at 0.5\,M$_\odot$.
\section{Conclusion}
\label{sec:conclusion}
We use solar neighbourhood stars to determine a new fiducial IMF within the framework of our local vertical Milky Way disc model (JJ-model).
 For that we carefully select volume complete samples based on dereddened and binary corrected Hipparcos and CNS\,5 data. Then we use Galaxia to create the corresponding mock observations from our JJ-model. We construct a likelihood by assuming a discrete Poisson process for the star count in magnitude bins differentiating between dwarfs and giants. With MCMC simulations we sample the PDF of the two slope IMF parameters. The derived IMF has a low mass power law index of $\alpha_1=1.49\pm 0.08$, a power law break at $\mathrm{m}_1=1.39\pm 0.05$\,M$_\odot$, a high mass index of $\alpha_2 = 3.02\pm 0.06$, a mass factor of $\mathrm{mf}=1.09\pm 0.04$ with respect to our paper\,I mass normalisation. Except for physics not accurately represented in our model (binaries, inhomogeneous ISM, variable stars, low mass stellar atmospheres, metal enrichment law, thick disc) these findings are robust in our observationally backed mass range from 0.5 to 10\,M$_\odot$. An empirically driven low mass extension adds $\alpha_0=1.26$ and $\mathrm{m}_0=0.5$ and decreases the mass factor to $1.03$.\\
Independently from us the Besan\c{c}on model using Tycho\,2 colour projections as observational constraints favours similar IMFs. The steep high mass slopes decrease the number of SN\,II ejected by an SSP compared to classical IMFs like Salpeter or Chabrier by a factor of about 3.\\
The future of analytic Galaxy modelling will see increasing modularity to incorporate up-to-date theoretical progress in stellar atmospheres and evolutionary tracks. Observational biases like binarity, selection effects or reddening will be accounted for from the models side and the observational samples will get diversified to overcome degeneracies in the various model ingredients like SFH, AVR and CEM. With ever more realistic Milky Way models more elusive concepts like the dark halo, the spiral arm structure or chemical yields will get further constraints to fit within the holistic framework.\\
To achieve that tools to measure the probability of model predictions given the data and schemes to optimise for the various data sets in an automated hierarchical fashion need to be implemented.\\ 
In the future we plan to implement a detailed chemical enrichment scheme into our disc model and exploit the elemental abundance information contained in surveys like RAVE, Gaia-ESO and SEGUE followed by the extension of the successful vertical description to different galactocentric radii.
\section*{Acknowledgements}
J. R. is funded by the DFG Research Centre SFB\,881 "The Milky Way System" through project A6. Further he is supported by the Heidelberg Graduate School of Fundamental Physics (HGSFP+).\\
We thank the anonymous referee for helpful advise on the manuscript.
\bibliography{paper}

\begin{thebibliography}{}

\bibitem[\protect\citeauthoryear{Aiyangar}{Aiyangar}{1988}]{Ai88}
Aiyangar S.,  1988, The lost notebook and other unpublished papers.
Narosa Pub. House, New Delhi

\bibitem[\protect\citeauthoryear{{Anderson} \& {Francis}}{{Anderson} \&
  {Francis}}{2012}]{an12}
{Anderson} E.,  {Francis} C.,  2012, Astronomy Letters, 38, 331

\bibitem[\protect\citeauthoryear{{Arenou}}{{Arenou}}{2011}]{Ar11}
{Arenou} F.,  2011 Vol.~1346 of American Institute of Physics Conference
  Series.
pp 107--121

\bibitem[\protect\citeauthoryear{{Aumer} \& {Binney}}{{Aumer} \&
  {Binney}}{2009}]{Au09}
{Aumer} M.,  {Binney} J.~J.,  2009, \mnras, 397, 1286

\bibitem[\protect\citeauthoryear{{Chabrier}}{{Chabrier}}{2003}]{Ch03}
{Chabrier} G.,  2003, \pasp, 115, 763

\bibitem[\protect\citeauthoryear{{Chen}, {Girardi}, {Bressan}, {Marigo},
  {Barbieri} \& {Kong}}{{Chen} et~al.}{2014}]{Ch14}
{Chen} Y.,  {Girardi} L.,  {Bressan} A.,  {Marigo} P.,  {Barbieri} M.,
  {Kong} X.,  2014, \mnras, 444, 2525

\bibitem[\protect\citeauthoryear{{Chiappini}, {Gratton} \& {R.}}{{Chiappini}
  et~al.}{1997}]{Ch97}
{Chiappini} C.,  {Gratton}   {R.} 1997, \apj, 477, 765

\bibitem[\protect\citeauthoryear{{Czekaj}, {Robin}, {Figueras}, {Luri} \&
  {Haywood}}{{Czekaj} et~al.}{2014}]{Cz14}
{Czekaj} M.~A.,  {Robin} A.~C.,  {Figueras} F.,  {Luri} X.,    {Haywood} M.,
  2014, \aap, 564, A102

\bibitem[\protect\citeauthoryear{{Dib}}{{Dib}}{2014}]{Di14}
{Dib} S.,  2014, \mnras, 444, 1957

\bibitem[\protect\citeauthoryear{{Dommanget} \& {Nys}}{{Dommanget} \&
  {Nys}}{2002}]{Do02}
{Dommanget} J.,  {Nys} O.,  2002, VizieR Online Data Catalog, 1274, 0

\bibitem[\protect\citeauthoryear{{ESA}}{{ESA}}{1997}]{Hi97}
{ESA} 1997, VizieR Online Data Catalog, 1239, 0

\bibitem[\protect\citeauthoryear{{Flynn}, {Holmberg}, {Portinari}, {Fuchs} \&
  {Jahrei{\ss}}}{{Flynn} et~al.}{2006}]{Fl06}
{Flynn} C.,  {Holmberg} J.,  {Portinari} L.,  {Fuchs} B.,    {Jahrei{\ss}} H.,
  2006, \mnras, 372, 1149

\bibitem[\protect\citeauthoryear{{Foreman-Mackey}, {Hogg}, {Lang} \&
  {Goodman}}{{Foreman-Mackey} et~al.}{2013}]{Fa13}
{Foreman-Mackey} D.,  {Hogg} D.~W.,  {Lang} D.,    {Goodman} J.,  2013, \pasp,
  125, 306

\bibitem[\protect\citeauthoryear{{Goodman} \& {Weare}}{{Goodman} \&
  {Weare}}{2010}]{Go10}
{Goodman} J.,  {Weare} J.,  2010, Communications in Applied Mathematics and
  Computational Science, 5

\bibitem[\protect\citeauthoryear{{Haywood}, {Robin} \& {Creze}}{{Haywood}
  et~al.}{1997}]{Ha97a}
{Haywood} M.,  {Robin} A.~C.,    {Creze} M.,  1997, \aap, 320, 428

\bibitem[\protect\citeauthoryear{{H{\o}g}, {Fabricius}, {Makarov}, {Urban},
  {Corbin}, {Wycoff}, {Bastian}, {Schwekendiek} \& {Wicenec}}{{H{\o}g}
  et~al.}{2000}]{Hoe00}
{H{\o}g} E.,  {Fabricius} C.,  {Makarov} V.~V.,  {Urban} S.,  {Corbin} T.,
  {Wycoff} G.,  {Bastian} U.,  {Schwekendiek} P.,    {Wicenec} A.,  2000, \aap,
  355, L27

\bibitem[\protect\citeauthoryear{{Holberg}, {Sion}, {Oswalt}, {McCook}, {Foran}
  \& {Subasavage}}{{Holberg} et~al.}{2008}]{Ho08}
{Holberg} J.~B.,  {Sion} E.~M.,  {Oswalt} T.,  {McCook} G.~P.,  {Foran} S.,
  {Subasavage} J.~P.,  2008, \aj, 135, 1225

\bibitem[\protect\citeauthoryear{{Holmberg} \& {Flynn}}{{Holmberg} \&
  {Flynn}}{2000}]{Ho00}
{Holmberg} J.,  {Flynn} C.,  2000, \mnras, 313, 209

\bibitem[\protect\citeauthoryear{{Holmberg}, {Flynn} \& {Lindegren}}{{Holmberg}
  et~al.}{1997}]{Ho97}
{Holmberg} J.,  {Flynn} C.,    {Lindegren} L.,  1997, in Hipparcos - Venice '97
  Vol.~402 of ESA Special Publication, {Towards an Improved Model of the
  Galaxy}

\bibitem[\protect\citeauthoryear{{Jahrei{\ss}} \& {Wielen}}{{Jahrei{\ss}} \&
  {Wielen}}{1997}]{Ja97}
{Jahrei{\ss}} H.,  {Wielen} R.,  1997, in Hipparcos - Venice '97 Vol.~402 of
  ESA Special Publication.
pp 675--680

\bibitem[\protect\citeauthoryear{{Just}, {Gao} \& {Vidrih}}{{Just}
  et~al.}{2011}]{Ju11}
{Just} A.,  {Gao} S.,    {Vidrih} S.,  2011, \mnras, 411, 2586

\bibitem[\protect\citeauthoryear{{Just} \& {Jahrei\ss}}{{Just} \&
  {Jahrei\ss}}{2010}]{JJ}
{Just} A.,  {Jahrei\ss} H.,  2010, \mnras, 402, 461

\bibitem[\protect\citeauthoryear{{Kroupa}, {Tout} \& {Gilmore}}{{Kroupa}
  et~al.}{1993}]{Kr93}
{Kroupa} P.,  {Tout} C.~A.,    {Gilmore} G.,  1993, \mnras, 262, 545

\bibitem[\protect\citeauthoryear{{Lallement}, {Vergely}, {Valette},
  {Puspitarini}, {Eyer} \& {Casagrande}}{{Lallement} et~al.}{2014}]{La14}
{Lallement} R.,  {Vergely} J.-L.,  {Valette} B.,  {Puspitarini} L.,  {Eyer} L.,
     {Casagrande} L.,  2014, \aap, 561, A91

\bibitem[\protect\citeauthoryear{{Lindegren}, {Mignard}, {S{\"o}derhjelm} \&
  {et.al.}}{{Lindegren} et~al.}{1997}]{Li97}
{Lindegren} L.,  {Mignard} F.,  {S{\"o}derhjelm} S.,    {et.al.} 1997, \aap,
  323, L53

\bibitem[\protect\citeauthoryear{{Ly}, {Malkan}, {Kashikawa}, {Shimasaku},
  {Doi}, {Nagao}, {Iye}, {Kodama}, {Morokuma} \& {Motohara}}{{Ly}
  et~al.}{2007}]{Ly07}
{Ly} C.,  {Malkan} M.~A.,  {Kashikawa} N.,  {Shimasaku} K.,  {Doi} M.,  {Nagao}
  T.,  {Iye} M.,  {Kodama} T.,  {Morokuma} T.,    {Motohara} K.,  2007, \apj,
  657, 738

\bibitem[\protect\citeauthoryear{{Marigo}, {Girardi}, {Bressan}, {Groenewegen},
  {Silva} \& {Granato}}{{Marigo} et~al.}{2008}]{Ma08}
{Marigo} P.,  {Girardi} L.,  {Bressan} A.,  {Groenewegen} M.~A.~T.,  {Silva}
  L.,    {Granato} G.~L.,  2008, \aap, 482, 883

\bibitem[\protect\citeauthoryear{{Mason}, {Wycoff}, {Hartkopf}, {Douglass} \&
  {Worley}}{{Mason} et~al.}{2001}]{Ma01}
{Mason} B.~D.,  {Wycoff} G.~L.,  {Hartkopf} W.~I.,  {Douglass} G.~G.,
  {Worley} C.~E.,  2001, \aj, 122, 3466

\bibitem[\protect\citeauthoryear{{Miller} \& {Scalo}}{{Miller} \&
  {Scalo}}{1979}]{Mi79}
{Miller} G.~E.,  {Scalo} J.~M.,  1979, \apjs, 41, 513

\bibitem[\protect\citeauthoryear{{Perryman}, {Lindegren}, {Kovalevsky} \&
  {et.al.}}{{Perryman} et~al.}{1997}]{Pe97}
{Perryman} M.~A.~C.,  {Lindegren} L.,  {Kovalevsky} J.,    {et.al.} 1997, \aap,
  323, L49

\bibitem[\protect\citeauthoryear{{Robin}, {Luri}, {Reyl{\'e}} \&
  {et.al.}}{{Robin} et~al.}{2012}]{Ro12}
{Robin} A.~C.,  {Luri} X.,  {Reyl{\'e}} C.,    {et.al.} 2012, \aap, 543, A100

\bibitem[\protect\citeauthoryear{{Robin}, {Reyl{\'e}}, {Derri{\`e}re} \&
  {Picaud}}{{Robin} et~al.}{2003}]{Ro03}
{Robin} A.~C.,  {Reyl{\'e}} C.,  {Derri{\`e}re} S.,    {Picaud} S.,  2003,
  \aap, 409, 523

\bibitem[\protect\citeauthoryear{{Salpeter}}{{Salpeter}}{1955}]{Sa55}
{Salpeter} E.~E.,  1955, \apj, 121, 161

\bibitem[\protect\citeauthoryear{{Scalo}}{{Scalo}}{1986}]{Sc86}
{Scalo} J.~M.,  1986, \fcp, 11, 1

\bibitem[\protect\citeauthoryear{{Schlafly}, {Green}, {Finkbeiner}, {Rix},
  {Bell}, {Burgett}, {Chambers}, {Draper}, {Hodapp}, {Kaiser}, {Magnier},
  {Martin}, {Metcalfe}, {Price} \& {Tonry}}{{Schlafly} et~al.}{2014}]{Sc14}
{Schlafly} E.~F.,  {Green} G.,  {Finkbeiner} D.~P.,  {Rix} H.-W.,  {Bell}
  E.~F.,  {Burgett} W.~S.,  {Chambers} K.~C.,  {Draper} P.~W.,  {Hodapp} K.~W.,
   {Kaiser} N.,  {Magnier} E.~A.,  {Martin} N.~F.,  {Metcalfe} N.,  {Price}
  P.~A.,    {Tonry} J.~L.,  2014, \apj, 786, 29

\bibitem[\protect\citeauthoryear{{Schmidt}}{{Schmidt}}{1959}]{Sc59}
{Schmidt} M.,  1959, \apj, 129, 243

\bibitem[\protect\citeauthoryear{{Sharma}, {Bland-Hawthorn}, {Johnston} \&
  {Binney}}{{Sharma} et~al.}{2011}]{Sh11}
{Sharma} S.,  {Bland-Hawthorn} J.,  {Johnston} K.~V.,    {Binney} J.,  2011,
  \apj, 730, 3

\bibitem[\protect\citeauthoryear{{Sion}, {Holberg}, {Oswalt}, {McCook},
  {Wasatonic} \& {Myszka}}{{Sion} et~al.}{2014}]{Si14}
{Sion} E.~M.,  {Holberg} J.~B.,  {Oswalt} T.~D.,  {McCook} G.~P.,  {Wasatonic}
  R.,    {Myszka} J.,  2014, \aj, 147, 129

\bibitem[\protect\citeauthoryear{{van den Bergh} \& {Sher}}{{van den Bergh} \&
  {Sher}}{1960}]{Va60}
{van den Bergh} S.,  {Sher} D.,  1960, Publications of the David Dunlap
  Observatory, 2, 203

\bibitem[\protect\citeauthoryear{{van Leeuwen}}{{van Leeuwen}}{2007}]{le07}
{van Leeuwen} F.,  2007, \aap, 474, 653

\bibitem[\protect\citeauthoryear{{Vergely}, {Ferrero}, {Egret} \&
  {Koeppen}}{{Vergely} et~al.}{1998}]{Ve98}
{Vergely} J.-L.,  {Ferrero} R.~F.,  {Egret} D.,    {Koeppen} J.,  1998, \aap,
  340, 543

\bibitem[\protect\citeauthoryear{{Weidner} \& {Kroupa}}{{Weidner} \&
  {Kroupa}}{2005}]{We05}
{Weidner} C.,  {Kroupa} P.,  2005, \apj, 625, 754

\bibitem[\protect\citeauthoryear{{Wielen}}{{Wielen}}{1974}]{Wi74}
{Wielen} R.,  1974, Highlights of Astronomy, 3, 395

\end{thebibliography}
\label{lastpage}
\end{document}